\documentclass[onecolumn]{article}
\usepackage{lineno}
\usepackage{algorithm}
\usepackage{algpseudocode}
\usepackage{mathtools}
\usepackage{tikz}
\usepackage{array}
\usepackage{booktabs}
\usepackage{makecell} 
\usepackage{upgreek}
\usepackage{graphicx,amsmath,cite,booktabs}

\begin{document}
\title{Preselection-Free Fiber-Optic Weak Measurement Sensing Framework with High-sensitivity}
\author{Zifu Su$^1$, Weiqian Zhao$^1$, Wanshou Sun$^1$, Hexiang Li$^1$, \\Yafei Yu$^{1,3^{*}}$, and Jindong Wang$^{2,4^{**}}$}
\maketitle 
\begin{abstract}    
    A preselection-free fiber-optic weak measurement sensing framework is proposed and experimentally verified in this paper. In view of the limitation that fiber-optic weak measurement require specific preselection, this scheme innovates theoretically and achieves high sensitivity sensing by optimizing the post-selection when single-mode optical fiber is used to generate random polarization state. The experimental results show that the phase sensing sensitivity of the sensor framework is $62 dB/rad$ (resolution of $1.6\times 10^{-5}rad$), the sensitivity for pressure sensing is $2.348 dB/N$ (resolution of $4.2\times 10^{-4} N$), and the sensitivity for temperature sensing is $12.695 dB/^{\circ} C$ (resolution of $7.8\times 10^{-5}\ ^{\circ} C$). The sensing performance is two to three orders of magnitude higher than that of traditional optical fiber sensing technology.
  \end{abstract}
  \bigskip            

\maketitle

\section{Introduction}

%
%

Fiber-optic sensing technology has garnered significant traction in industrial monitoring, military defense, and biomedical applications due to its inherent advantages: electromagnetic immunity, sub-nano-strain sensitivity, and distributed measurement capabilities \cite{FST1, FST2, FST3, FST4, FST5}. In particular, polarization-maintaining fiber (PMF) exhibit exceptional performance in polarization-sensitive sensing owing to their intrinsic birefringence. Recent advances in high-sensitivity pressure and temperature sensing have leveraged fiber Bragg grating (FBG)  \cite{RI_TS, HS_TS, Lateral_Pressure_Sen}, interferometer configuration  \cite{Lateral_Force_Sen, Enhenc_Sens_tem, Temp_Sens_Inter, Selectively_toluene}, and microstructured fibers \cite{Simulta_strain, Micro-structured}. Nevertheless, these systems suffer from three critical limitations:Architectural complexity requiring precise alignment of dual PMF or hybrid interferometer. Prohibitive costs associated with high-resolution spectrometers and specialized fabrication.Scalability challenges for industrial deployment.Consequently, the development of simplified, cost-effective, and ultra-sensitive fiber-optic sensing platforms remains imperative to meet escalating performance demands.
	
Quantum weak measurement (WM), introduced by Aharonov et al. in 1988 \cite{WM_first}, enables parameter amplification via weak value amplification (WVA) through optimized pre- and post-selection processes. 
The principle of WVA can be explained by the interference of probability amplitude \cite{WM_sense} in the context of quantum mechanics, or the interference of classical waves \cite{Interf_WM}. 
While WM has been extensively utilized in free-space metrology to achieve attometer-level displacement resolution \cite{Obser_SP, Noise_mitigat, Reali_ultra_small}, its integration into fiber-optic sensing remains nascent \cite{OUR_OWN}. Luo et al. \cite{Low_freq_hydr} demonstrated a WM-based hydrophone using $0.8m$ PMF and polycarbonate substrates, achieving $0.1Hz$ to $100Hz$ low-frequency response with $40dB$ noise suppression compared to Michelson interferometers. Liu et al. \cite{Liu:22} developed a polarization-encoded WM system exhibiting flat frequency response $0.1Hz$ to $10kHz$ and two-order noise reduction relative to conventional interferometric sensors.
	
However, the current optical fiber sensing system based on WVA technology has high requirements for optical fibers. It requires two polarization-preserving optical fibers with the same length and perpendicular alignment of fast and slow axes. In addition, the polarization state of incident light needs to be considered, which makes these optical fiber sensing systems more complex, so its application prospect is limited.
	
Here we report a preselection-free fiber-optic weak measurement sensing framework with high-sensitivity, which is simple in structure and highly sensitive. The proposed fiber-optic weak measurement sensing framework mainly consists of a polarization controller (PC), a linear polarizer, and a polarization-maintaining fiber (PMF). By optimizing the angle after weak measurement, the sensing system can achieve high sensitivity without complex hardware support or sophisticated demodulation methods. Experimental results show that the phase sensing sensitivity of the sensor framework is 62 dB/rad, the sensitivity for pressure sensing is $2.348 dB/N$, and the sensitivity for temperature sensing is $12.695 dB/^{\circ} C$. Compared with traditional optical fiber sensing methods, the sensitivity of this sensing scheme can be two to three orders of magnitude higher.
	
The remainder of this paper is organized as follows: Section 2 introduces the theory of polarization-based fiber-optic WM sensing without specific pre-selection, particularly how to improve fiber optic sensing sensitivity by optimizing the post-selection angle. Section 3 presents a simple structure for a fiber-optic WM sensing scheme without specific pre-selection and discusses the phase sensing, pressure sensing, and temperature sensing characteristics of the proposed fiber optic sensing scheme. Section 4 summarizes the proposed sensing scheme.
	
\section{Theoretical Framework}
Quantum WM schemes can typically be dividedinto four steps \cite{XU2024100518}, the first step is pre-selection, preparing the initial quantum state of the system; the second step is the weak coupling phase, where the system to be measured interacts weakly with the probe pointer; the third step is post-selection, projecting the evolved state of the composite system onto a predefined post-selection quantum state through projection measurement; the fourth step is data readout, observing the final state of the pointer via detection devices to indirectly obtain information about the target physical quantity of the system.
	
Considering the fiber-optic WM sensing scheme without specific pre-selection, the pre- and post-selection of quantum states can be regarded as the polarization of photons. The system pre-selection state is:
\begin{equation}
		|\varphi_{i} \rangle =\cos \theta e^{i\alpha}|H\rangle +\sin \theta e^{-i\alpha}|V\rangle,
\end{equation} 

where $\theta$ is the pre-selection angle of the pre-selected state, the pre-selection state represents the incidence of arbitrary polarized light. When the pre-selected state of light is transmitted through the optical fiber, it will be disturbed by random noise, and finally form a mixed state. The mixed state can be expressed as:
\begin{equation}
	\rho =p_{0}\frac{I}{2} +\left( 1-p_{0} \right) |\varphi_{i} \rangle \langle \varphi_{i} |.
\end{equation} 
	
The initial state of the probe is the initial spectral distribution of the light source, that is $|\phi \rangle =\int d\omega f\left( \omega \right) |\omega \rangle$, where $f\left( \omega \right) =\left( \pi \sigma^{2} \right)^{-\frac{1}{4}} e^{-\frac{\left( \omega -\omega_{0} \right)^{2}}{2\sigma^{2}}}$, $\sigma^{2} =<\omega^{2} >-<\omega >^{2}$ for spectral bandwidth, $\omega_{0}$ is the average frequency of the initial spectrum. The system and the measuring pointer have weak interaction operators: $\hat{U} =e^{-i\omega \tau \hat{A}}$, where $\tau$ is the delay generated internally by the sensing optical fiber PMF, $\hat{A} =|H\rangle \langle H|-|V\rangle \langle V|$is a system observable measure. The constraint of $\left| \sigma \tau \right| <<1$ is imposed to ensure operation within the weak coupling regime.
	
The weak coupling between the mixed state of the quantum system and the initial state of the pointer leads to the entangled joint state:
	\begin{equation}
		\rho_{joint} =\hat{U} \left( \rho \otimes |\phi \rangle \langle \phi | \right) \hat{U}^{\dag}
	\end{equation}
	
Project the entangled joint state onto the preset post-selection state: 
	\begin{equation}
		|\varphi_{f} \rangle =\left( e^{-i\epsilon}|H\rangle +e^{i\epsilon}|V\rangle \right)/\sqrt{2}, 
	\end{equation} 
	where $\epsilon$ is the post-selection angle. The output light power is given by the following formula:
	\begin{equation}
		\begin{aligned}
			I_{out}
			&=I_{in}\langle \phi_{f} |\rho_{joint} |\phi_{f} \rangle\\
			&= \left( 1-p_{0} \right) I_{in}\int \left[ \frac{1}{2} +\frac{1}{2} \sin 2\theta \cos 2\left( \omega \tau -\alpha -\epsilon \right) \right] f^{2}\left( \omega \right) d\omega\\&+\frac{p_{0}}{2} I_{in}\int f^{2}\left( \omega \right) d\omega\\ 
			&\approx \left( 1-p_{0} \right) I_{in}\left[ \frac{1}{2} +\frac{1}{2} \sin 2\theta \cos 2\left( \omega \tau -\alpha -\epsilon \right) \right]+\frac{p_{0}}{2} I_{in},
		\end{aligned}
	\end{equation}
	here, $I_{in}$ is the incident light intensity before the subsequent selection step. The amount of delay $2\tau$ generated inside the sensing optical fiber PMF can be written as: $2\tau =2\tau_{0} +2\bigtriangleup \tau$, $\tau_{0} >>\bigtriangleup \tau$, where $2\tau_0$ is the initial delay quantity inside the sensing optical fiber PMF, $2\Delta\tau$ is the delay of the small change of sensing measurement by the sensing optical fiber PMF.
	
	First of all, when the optical fiber sensing delay $\Delta \tau=0$,by scanning the post-selection angle $\epsilon$, when $2\omega_0\tau-2\alpha-2\epsilon=k\pi$ , where $k$ is odd or even, the output light intensity will be minimized, and the minimum value of the output beam intensity obtained $I_{min}$ is:
	\begin{equation}
		I_{min}=\left( 1-p_{0} \right) I_{in}\left( \frac{1}{2} -\frac{1}{2} \left| \sin 2\theta \right| \right) +\frac{p_{0}}{2} I_{in},
	\end{equation} 
	
	Secondly, on the basis of the post-selected angle $\epsilon$, the post-selected angle is further changed to $\epsilon+\Delta\epsilon$, and the changed $\Delta\epsilon$ is marked as the optimized post-selected angle. The output light intensity obtained by selecting the optimized post-selected angle is $I_0$,
	\begin{equation}
		I_{0}=\left( 1-p_{0} \right) I_{in}\left( \frac{1}{2} -\frac{1}{2} \left| \sin 2\theta \right| +\left| \sin 2\theta \right| \sin^{2} \Delta \epsilon \right) +\frac{p_{0}}{2} I_{in}.
	\end{equation} 
	
	Finally, the sensing is carried out under the selected optimized post-selection angle $\Delta \epsilon$. External disturbance causes the sensing optical fiber PMF to produce a small delay $2\Delta \tau$, the resulting output light intensity is denoted by $I_{1}$,
	\begin{equation}
		I_{1}=\left( 1-p_{0} \right) I_{in}\left( \frac{1}{2} -\frac{1}{2} \left| \sin 2\theta \right| +\left| \sin 2\theta \right| \sin^{2} \left( \omega_{0} \Delta \tau -\Delta \epsilon \right) \right) +\frac{p_{0}}{2} I_{in}
	\end{equation} 
	
	The resulting intensity contrast is given by the following formula:
	\begin{equation}
		\eta =\frac{I_{0}-I_{1}}{I_{0}-I_{min}} \approx \frac{2\omega_{0} \Delta \tau}{\Delta \epsilon} ,\  \omega_{0} \Delta \tau <<\Delta \epsilon <<1.
	\end{equation}

	The obtained intensity contrast $\eta$ is independent of the pre-selection state of the incident light and $p_{0}$, so the sensing method does not need to pre-select the polarization state. The phase difference of optical fiber sensing measurement can be obtained by the intensity contrast as $2\omega_{0} \Delta \tau \approx \eta \Delta \epsilon$, the phase difference resolution of the sensor is only related to the size of our optimized post-selection angle $\Delta\epsilon$ and the minimum output light intensity change that the power meter can distinguish. When $\theta =\pm 45^{\circ}$, $p_{0}=0$ the change of output light intensity $\Delta I=10lg\frac{I_{1}}{I_{in}} -10lg\frac{I_{0}}{I_{in}} \approx 10lg\left( 1-\eta \right)$. Due to the minimum resolution of commercial optical power meters typically being 0.001 dB, this means that the minimum resolution for intensity contrast in our sensing scheme is 0.00023. Under the same meter measurement, the optimized selection angle amplifies the intensity contrast, and the smaller the optimized post-selection angle, the greater the amplification effect, allowing for a narrower phase range to be sensed. Due to $\left| \sigma \tau \right| <<1$, the use of narrow band light source can ensure that the sensing optical fiber remains highly sensitive over a long distance.
	
	\section{Experimental setup}
	Fig. 1 shows a simple structure of a preselection-free fiber-optic WM sensing experimental system. The light source used is a tunable single-frequency laser (DenseLight, laser box) operating at a central wavelength of $1550nm$ with a linewidth of $3.7kHz$. After the output polarization light of the light source passes through a single mode fiber (SMF), its polarization state changes randomly and is disturbed by random noise, and the mixed state state is obtained at this time.
	
	A polarization controller (PC) is connected to the SMF to fine-tune the input polarization. After the PC, the light enters a $30cm$ long polarization-maintaining fiber (PMF), where the fast and slow axes define the $H$ and $V$ polarization basis of the system. The PMF functions as the sensing module, where weak interactions are introduced by external perturbations, leading to polarization-dependent phase shifts.
	
	Post-selection is performed using a polarization control module (Thorlabs, FiberBench), which consists of a quarter-wave plate (QWP) and a high-extinction-ratio linear polarizer (P, extinction ratio: 10000:1). The QWP is oriented at $45^{\circ}$ relative to the slow axis of the PMF, while the polarizer P is aligned at $45^{\circ} + \epsilon$, where $\epsilon$ is the adjustable post-selection angle. The polarizer can be rotated to optimize the post-selection phase and achieve sensitivity tuning.
	
	Finally, the post-selected optical signal is measured by a power meter (Thorlabs, PM100D). When a small external disturbance is applied to the PMF, it induces a weak interaction that manifests as a phase shift between the $H$ and $V$ polarization components. This phase variation is translated into an intensity change in the output beam, as predicted by WM theory.
	
	\begin{figure}[htbp]
		\centering
			\includegraphics[width=1\linewidth]{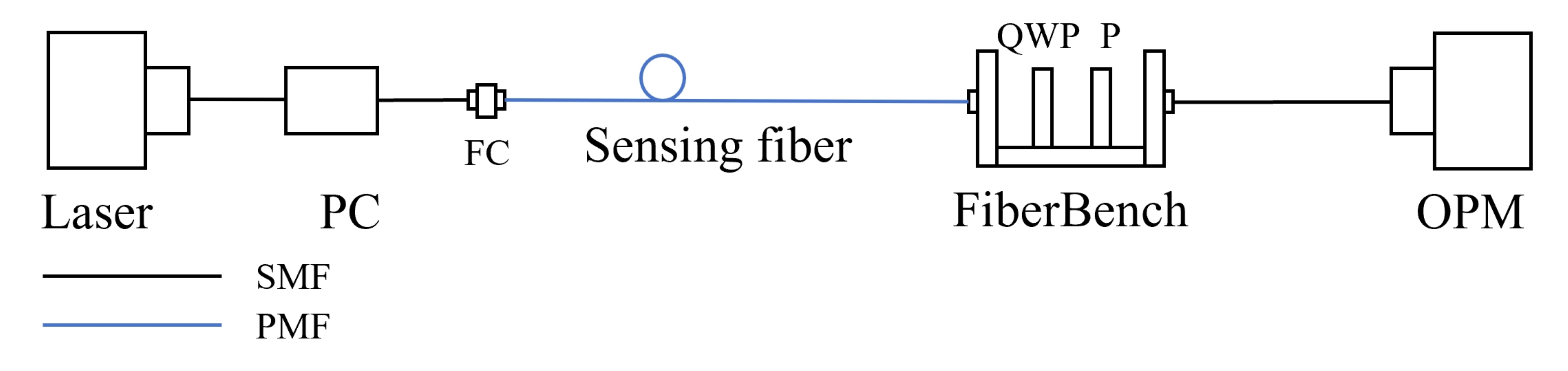}
		\caption{Preselection-free fiber-optic weak measurement sensing experimental system}
		\label{fig:1}
	\end{figure}

	\subsection{Phase sensing characteristics}
	In order to verify the feasibility of high sensitivity sensing in the preselection-free fiber-optic WM sensing experimental system shown in Fig. 1, we use the PC to select different preselected states into SMF, and verify the principle of optical fiber weak measurement phase sensing by using the different mixed states output from SMF.
	
	In the experimental process, we first adjusted the optical axis direction of QWP and P in the post-selection module. By adjusting only the post-selection angle $\epsilon$ , we obtained the minimum output light intensity $I_{min}$ and the pre-selection angle $\theta=20^{\circ}$ of the initial pre-selection state. Then, the optical axis direction of P is finely adjusted to set the optimized post-selection angle $\Delta \epsilon$. Since the input light is narrow band light, the simulated sensing phase $2\omega_{0} \Delta \tau$  can be simulated by rotating P. P rotated $0.25^{\circ}$ and the corresponding phase change $2\omega_{0} \Delta \tau$  is $8.73\times 10^{-3}rad$. The experimental results are shown in Fig. 2. In order to reduce the measurement error, each data point of output light intensity is recorded by a power meter for $60s$.

	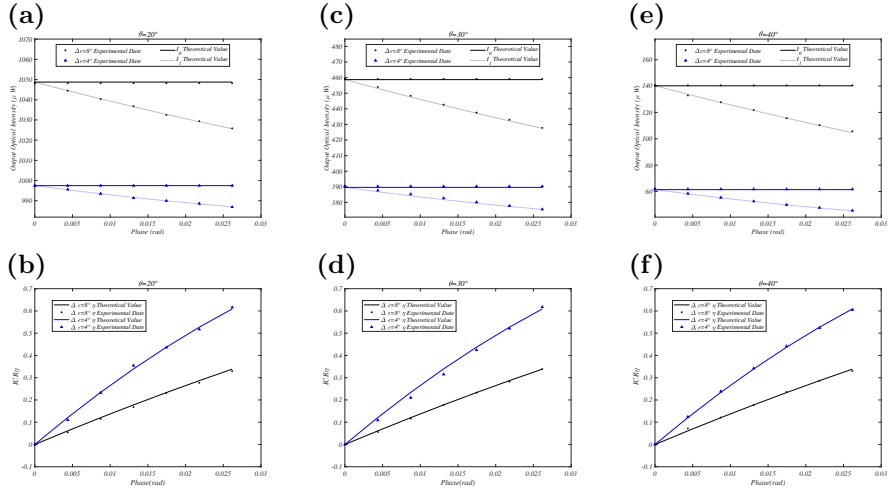
\begin{figure}[h]
		\centering
		
		\begin{minipage}{0.32\textwidth}
			\centering
			\begin{tikzpicture}
				\node[anchor=south west,inner sep=0] (image1) at (0,0) {\includegraphics[width=\linewidth]{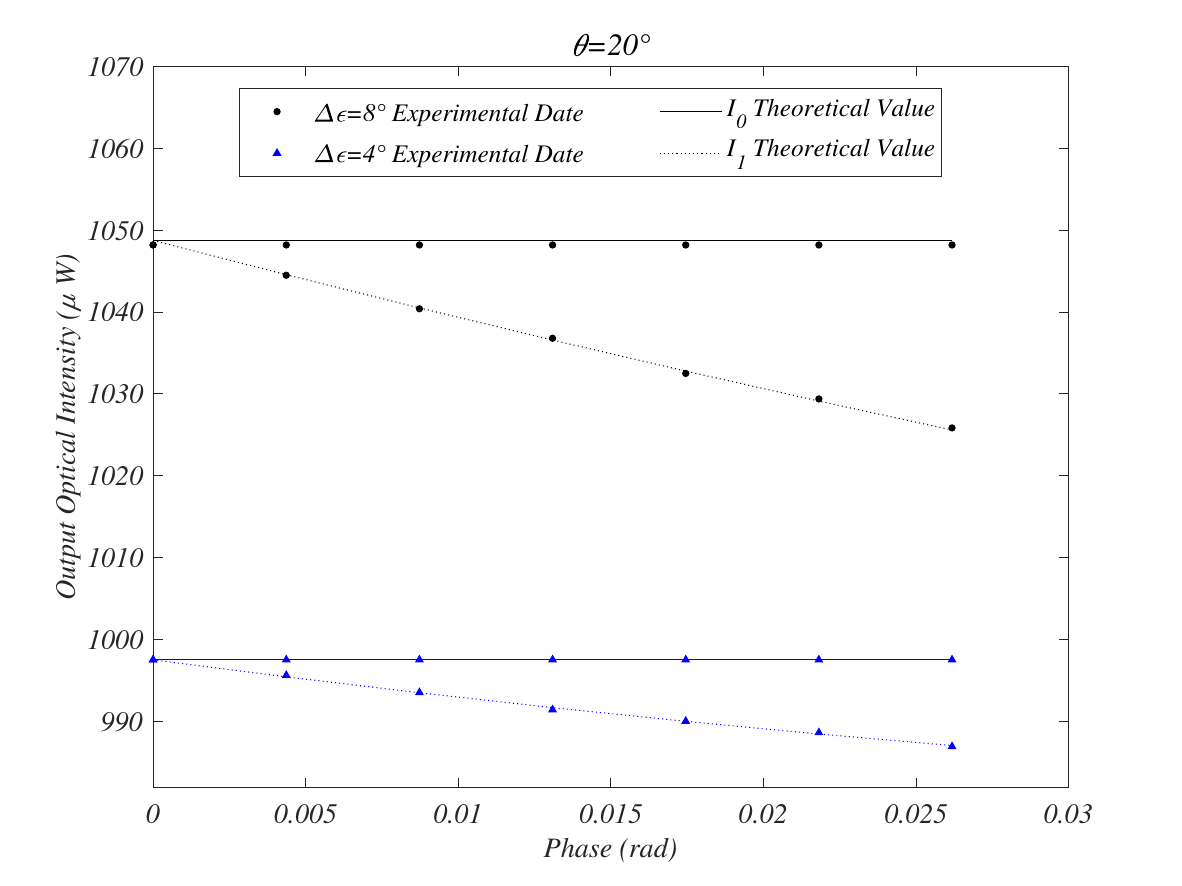}};
				\node at (0.40,3.0) {\textbf{(a)}};
			\end{tikzpicture}
		\end{minipage}
		\hfill
		\begin{minipage}{0.32\textwidth}
			\centering
			\begin{tikzpicture}
				\node[anchor=south west,inner sep=0] (image3) at (0,0) {\includegraphics[width=\linewidth]{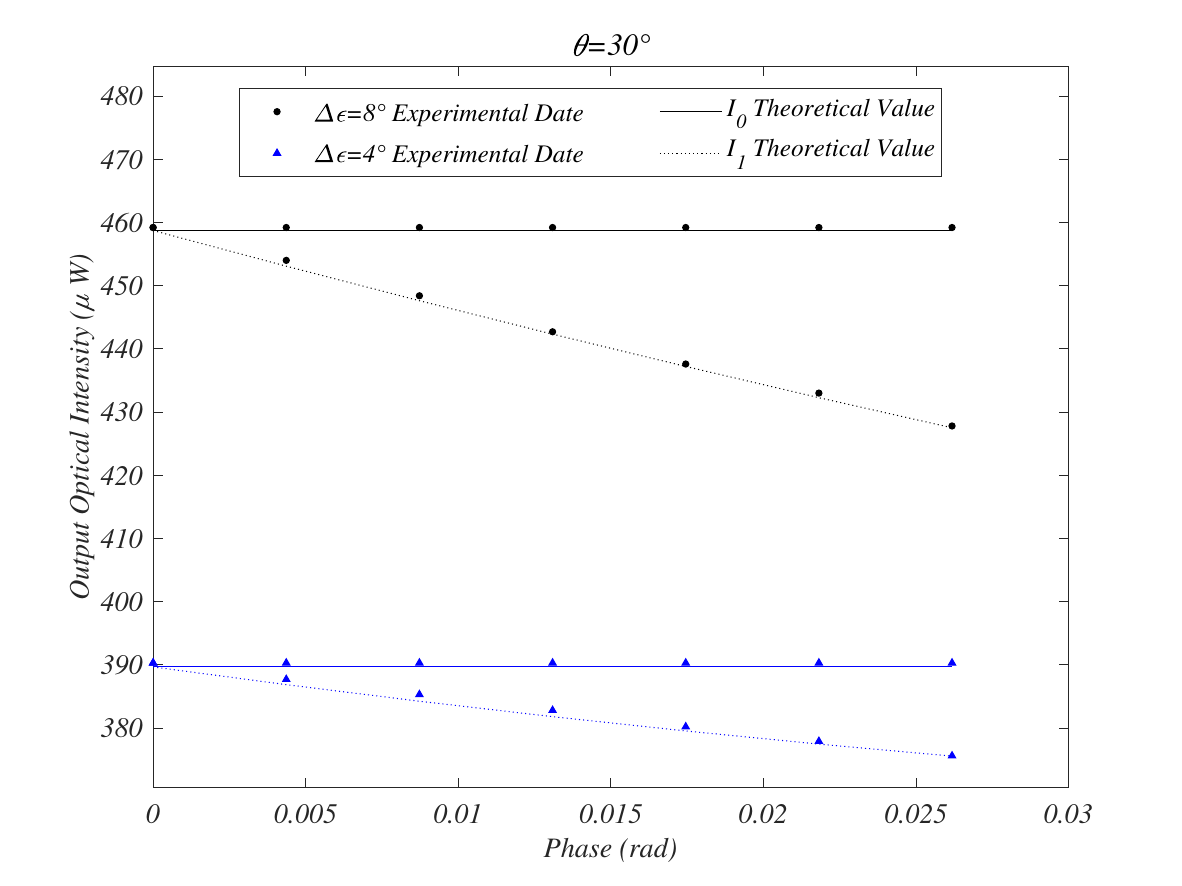}};
				\node at (0.40,3.0) {\textbf{(c)}};
			\end{tikzpicture}
		\end{minipage}
		\hfill
		\begin{minipage}{0.32\textwidth}
			\centering
			\begin{tikzpicture}
				\node[anchor=south west,inner sep=0] (image5) at (0,0) {\includegraphics[width=\linewidth]{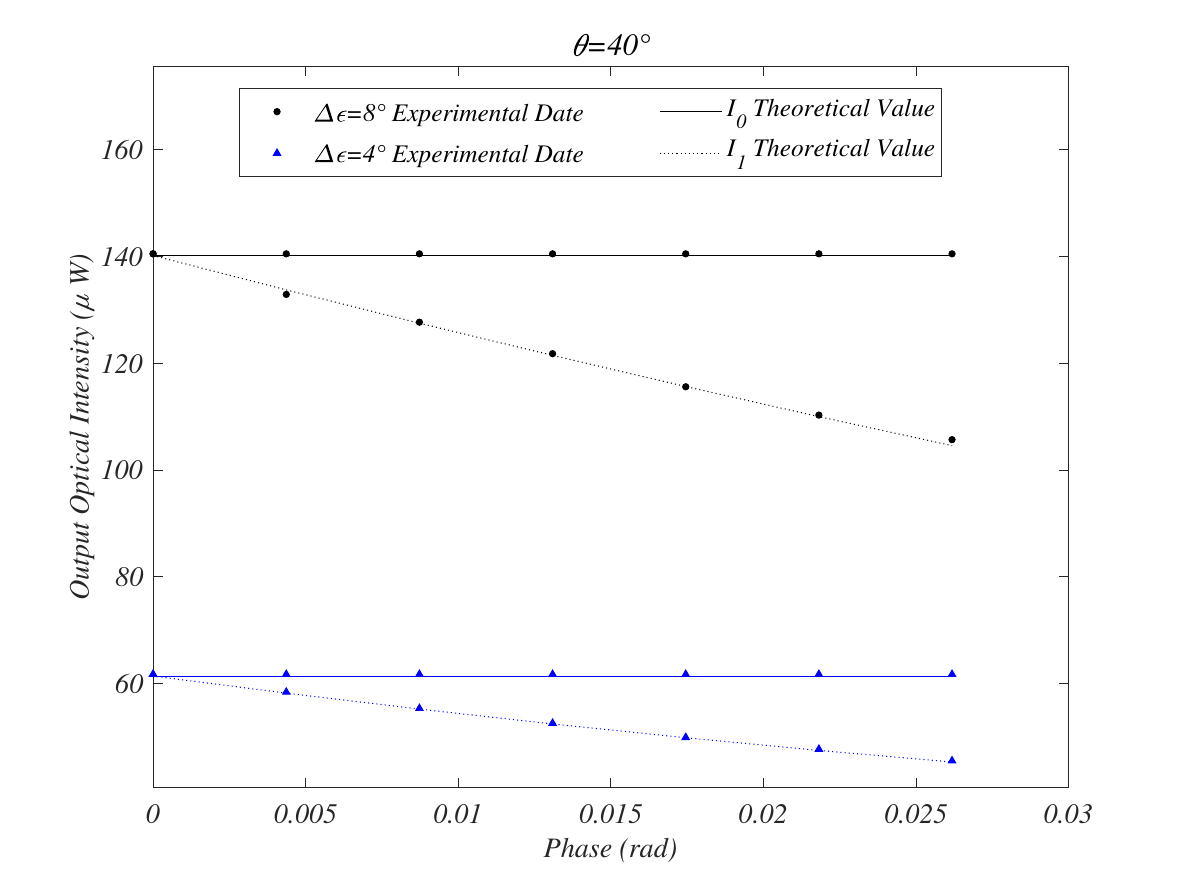}};
				\node at (0.40,3.0) {\textbf{(e)}};
			\end{tikzpicture}
		\end{minipage}
		
		\vspace{0cm}
		
		\begin{minipage}{0.32\textwidth}
			\centering
			\begin{tikzpicture}
				\node[anchor=south west,inner sep=0] (image2) at (0,0) {\includegraphics[width=\linewidth]{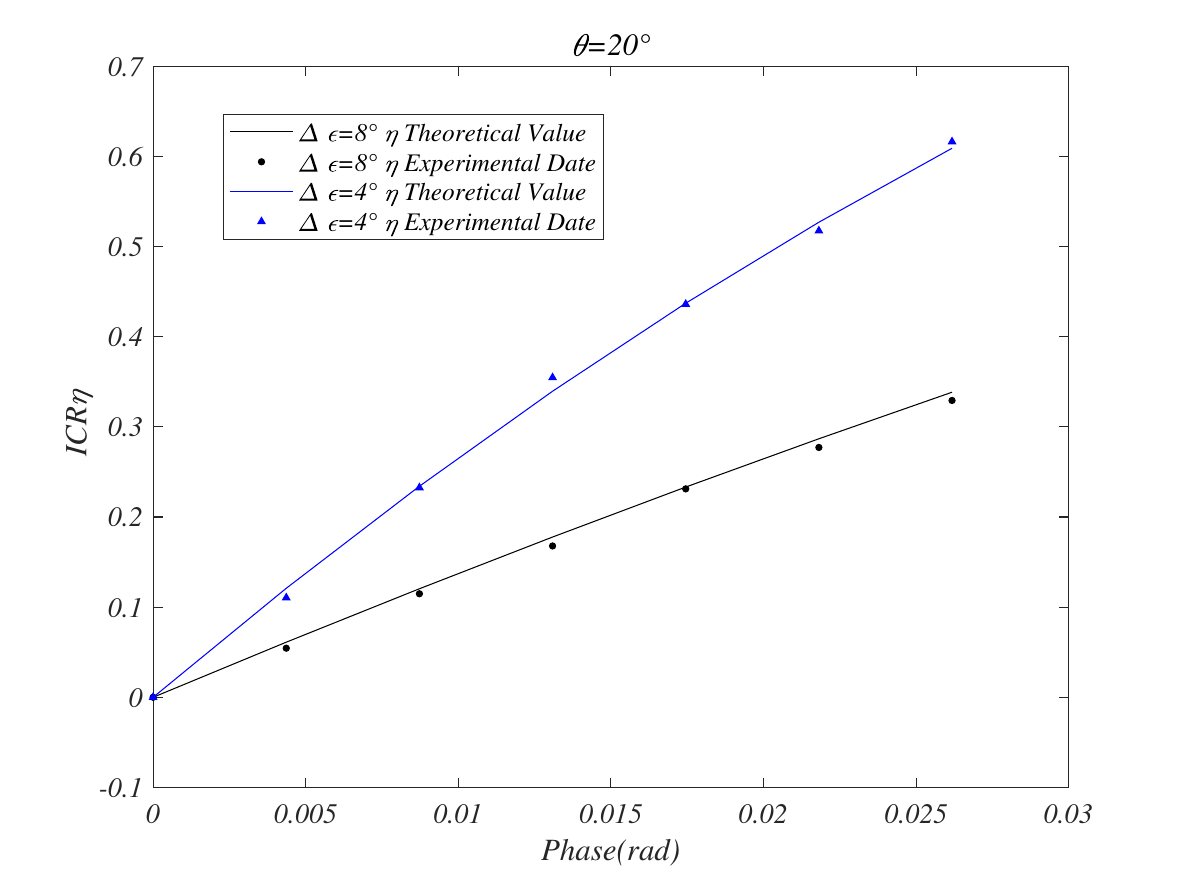}};
				\node at (0.40,3.0) {\textbf{(b)}};
			\end{tikzpicture}
		\end{minipage}
		\hfill
		\begin{minipage}{0.32\textwidth}
			\centering
			\begin{tikzpicture}
				\node[anchor=south west,inner sep=0] (image4) at (0,0) {\includegraphics[width=\linewidth]{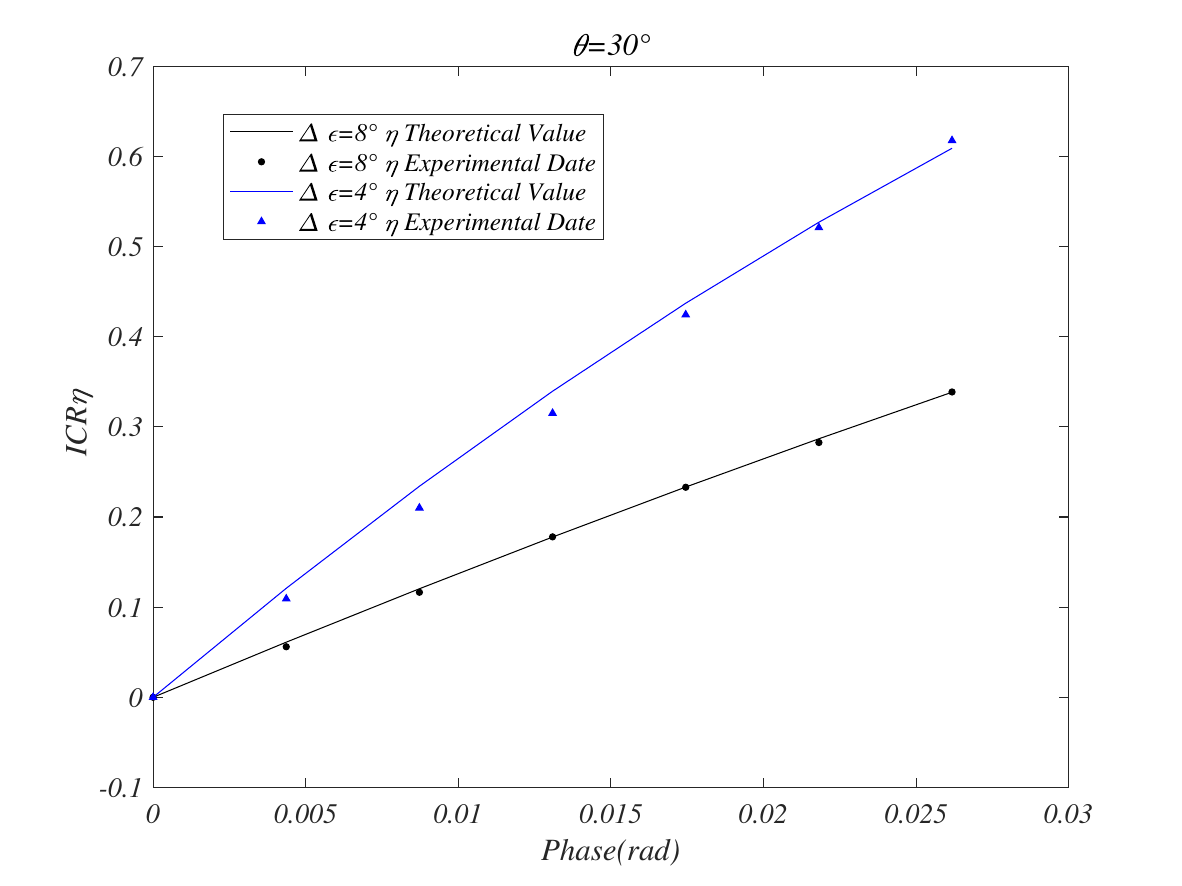}};
				\node at (0.40,3.0) {\textbf{(d)}};
			\end{tikzpicture}
		\end{minipage}
		\hfill
		\begin{minipage}{0.32\textwidth}
			\centering
			\begin{tikzpicture}
				\node[anchor=south west,inner sep=0] (image6) at (0,0) {\includegraphics[width=\linewidth]{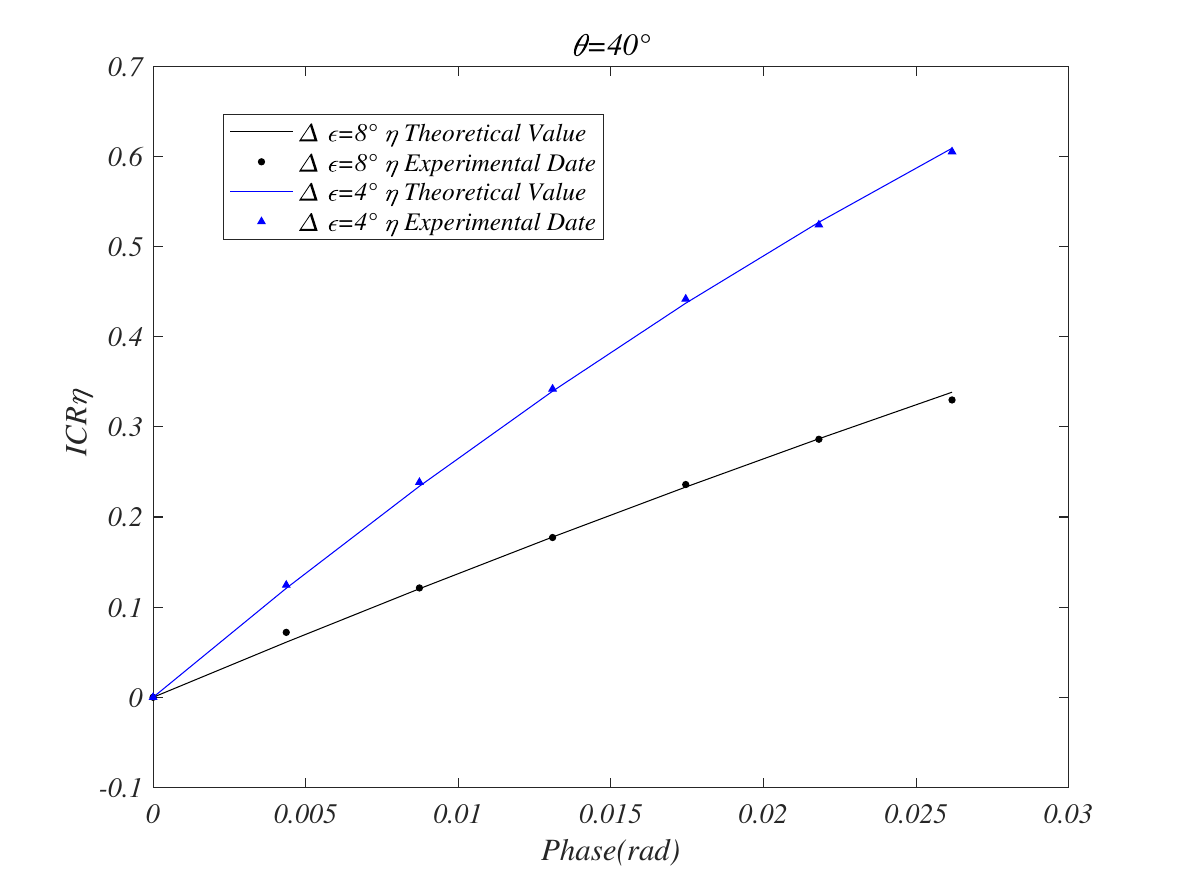}};
				\node at (0.40,3.0) {\textbf{(f)}};
			\end{tikzpicture}
		\end{minipage}
		
		\caption{Experimental results of phase sensing with pre-selection angles $\theta=20^{\circ}, 30^{\circ}, 40^{\circ}$. Figures (a), (c), and (e) show the theoretical and experimental responses of the selected optical power $I_1$ to the sensing phase $2\omega_0 \Delta \tau$ for the optimized post-selection angles $\Delta\epsilon$ of $4^{\circ}$ and $8^{\circ}$. Figures (b), (d), and (f) show the corresponding intensity contrast $\eta$ versus the sensing phase under the same post-selection conditions}
		\label{fig:2}
	\end{figure}

	In phase-sensing experiments with a pre-selected angle $\theta =20^{\circ}$, the optimized post-selection angle $\Delta\epsilon$ is configured at $4^{\circ}$ (blue) and $8^{\circ}$ (black), as illustrated in Fig. 2(a). The solid curve corresponds to the baseline optical power $I_0$ under non-interferometric conditions, while the dashed curve quantifies the functional dependence of the post-selected optical power $I_1$ on the phase variation $2\omega_{0} \Delta \tau$ .Obviously, the smaller optimized post-selection angle  $\Delta\epsilon$ leads to lower post-selection optical power $I_0$ and $I_1$, which is consistent with the theoretical prediction of post-selection probability. 
	
	Fig. 2(b) shows the intensity contrast $\eta$ response to phase variation  $2\omega_{0} \Delta \tau$ under different optimized post-selection angle  $\Delta\epsilon$, with a pre-selected angle fixed at $\theta=20^{\circ}$. The solid line represents the theoretical simulation results under the condition $\omega_{0} \Delta \tau <<\Delta \epsilon <<1$ and $\sigma^{2} \tau^{2} <<1$, showing a linear relationship of $\eta \approx 2\omega_{0} \Delta \tau /\Delta \epsilon$, which is very consistent with the experimental results.The intensity contrast $\eta$ is proportional to the sensing phase $2\omega_{0} \Delta \tau$ and inversely proportional to the optimized post-selection angle $\Delta\epsilon$. The smaller optimized post-selection angle will make the intensity contrast $\eta$ response of each unit sensing phase  $2\omega_{0} \Delta \tau$ larger, so as to improve the sensitivity of the detection phase.
	
	By adjusting the PC to modify the pre-selection state, the optical axes of the QWP and P in the post-selection module are subsequently realigned to achieve a new minimized output intensity $I_{min}$, thereby generating a renewed input pre-selection state. The phase sensing experiment was repeated with the pre-selection angle $\theta =20^{\circ}$, and the phase sensing experiment was continued with the pre-selection angle $\theta =30^{\circ}$ [Figs. 2 (c) and (d)] and $\theta =40^{\circ}$ [Figs. 2 (e) and (f)]. The experimental results are shown in Fig. 2.
	
	The experimental phase sensing results across varying pre-selection angles [Figs. 2(b), (d), (f)] demonstrate remarkable agreement with the theoretical framework of WM sensing without specific pre-selection. Quantitative analysis confirms the intensity contrast follows the predicted relationship: $\eta \approx 2\omega_{0} \Delta \tau /\Delta \epsilon$. When $\Delta \epsilon =4^{\circ}$, the obtained phase sensitivity is $14.28dB/rad$. The measured phase sensitivity corresponds to $62dB/rad$ when expressed in intensity-equivalent units, and the weak measurement phase sensitivity is quantified. The sensitivity-phase relationship is strictly controlled by $\Delta \epsilon$  and has nothing to do with the pre-selected state of the input. Even if the pre-selection state changes randomly, the sensing system can still maintain high sensitivity, which verifies the feasibility of achieving preselection-free fiber-optic weak measurement sensing with high sensitivity.
	
	\subsection{Pressure sensing characteristics}
	In order to study the physical sensing mechanism of the preselection-free fiber-optic weak measurement sensing system with high sensitivity, lateral pressure stimulation was systematically applied to PMF, and its pressure sensing characteristics were tested and analyzed. When the sensing optical fiber PMF is affected by mechanical load changes, its internal structure and fiber length change, leading to variations in the birefringence difference between the fast and slow axes of the fiber. This introduces changes in the optical phase difference between $H$ and $V$ polarizations, which can be described by the following formula:
  \begin{equation}
2\omega_0\Delta\tau =
\frac{2\pi}{\lambda_0}
\left(L\frac{\partial B}{\partial F}
      +B\frac{\partial L}{\partial F}\right)\Delta F,
\end{equation}  
	where $B$ is the difference of refractive index between fast and slow axes of optical fiber, $L$ represents the length of PMF, $\Delta F$ is the lateral pressure change of PMF, and $\frac{\partial L}{\partial F}$  describes the fiber elongation caused by the change of mechanical load of PMF \cite{Zu_2011} .In the absence of external interference, the phase difference produced by a certain length of PMF is generally considered proportional to the applied lateral pressure. The final relationship between the change in phase difference and lateral pressure is: $2\omega_{0} \Delta \tau \sim 2\uppi \Delta F/\lambda_{0}$ . Thus, the changes in the phase difference of the sensing fiber, as measured by WM sensing theory, reflect the variations in the lateral pressure on the sensing fiber.
	
	The experimental setup for the preselection-free fiber-optic weak measurement pressure sensing is shown in Fig. 3. In the experiment, a type of optical fiber and a $30cm$ long PMF sensing fiber are placed between two flat glass plates. The load length of the sensing fiber is $38 mm$, and then individual weights of $10g$ are sequentially pressed onto the top of the flat glass plates. The sensing fiber PMF experiences a pressure increment of $0.05 N$. The PC prepares different pre-selection to verify that the sensing experiment is independent of the input pre-selection state. For the given optimized post-selection angle, the experimental value of the output light intensity after post-selection varies with the sensing pressure. The power meter monitors the selected output light intensity and detects the magnitude of the applied pressure through the power of the output light.
	\begin{figure}[htbp]
		\centering
		\includegraphics[width=1\linewidth]{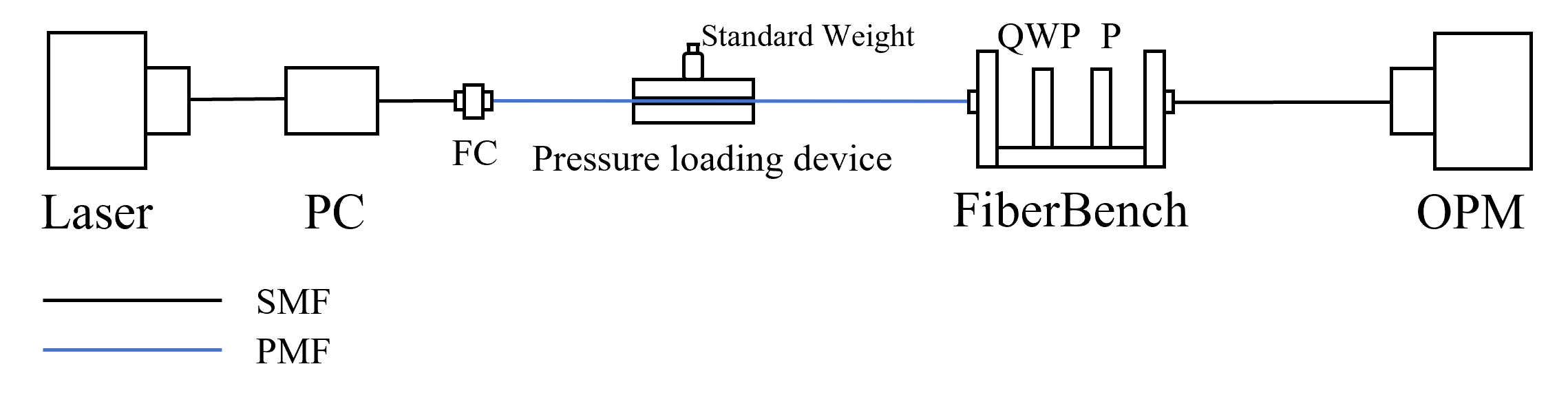}
		\caption{Schematic diagram of the preselection-free fiber-optic weak measurement pressure sensing experiment}
		\label{fig:4}
	\end{figure}
	
	The experimental results are shown in Fig 4. The selected optimized post-selection angles $\Delta\epsilon$  were $4^{\circ}$ and $8^{\circ}$, and three different pre-selection states were input. Pressure sensing experiments were carried out at different values of pre-selection angle $\theta=20^{\circ}$ [Fig.4 (a) and (b)], $30^{\circ}$ [Fig. 4 (c) and (d)] and $40^{\circ}$ [Fig. 4 (e) and (f)].
	
	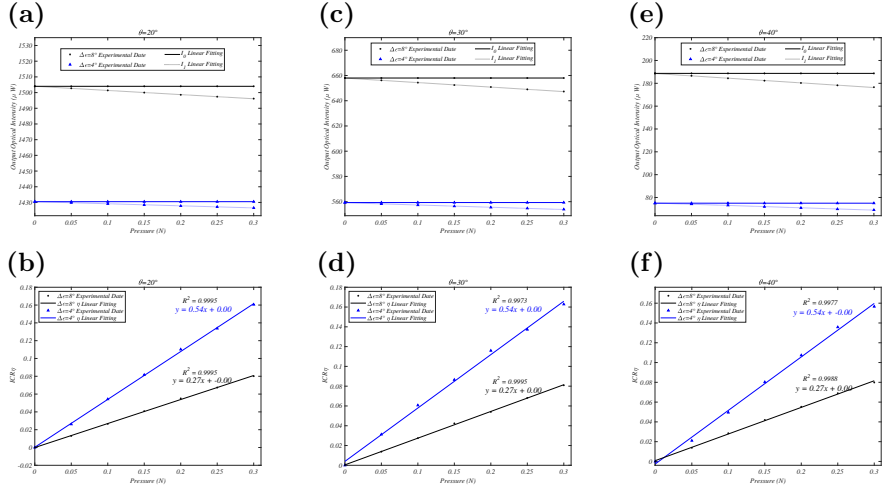
\begin{figure}[h]
		\centering
		
		\begin{minipage}{0.32\textwidth}
			\centering
			\begin{tikzpicture}
				\node[anchor=south west, inner sep=0] (image1) at (0,0) {\includegraphics[width=\linewidth]{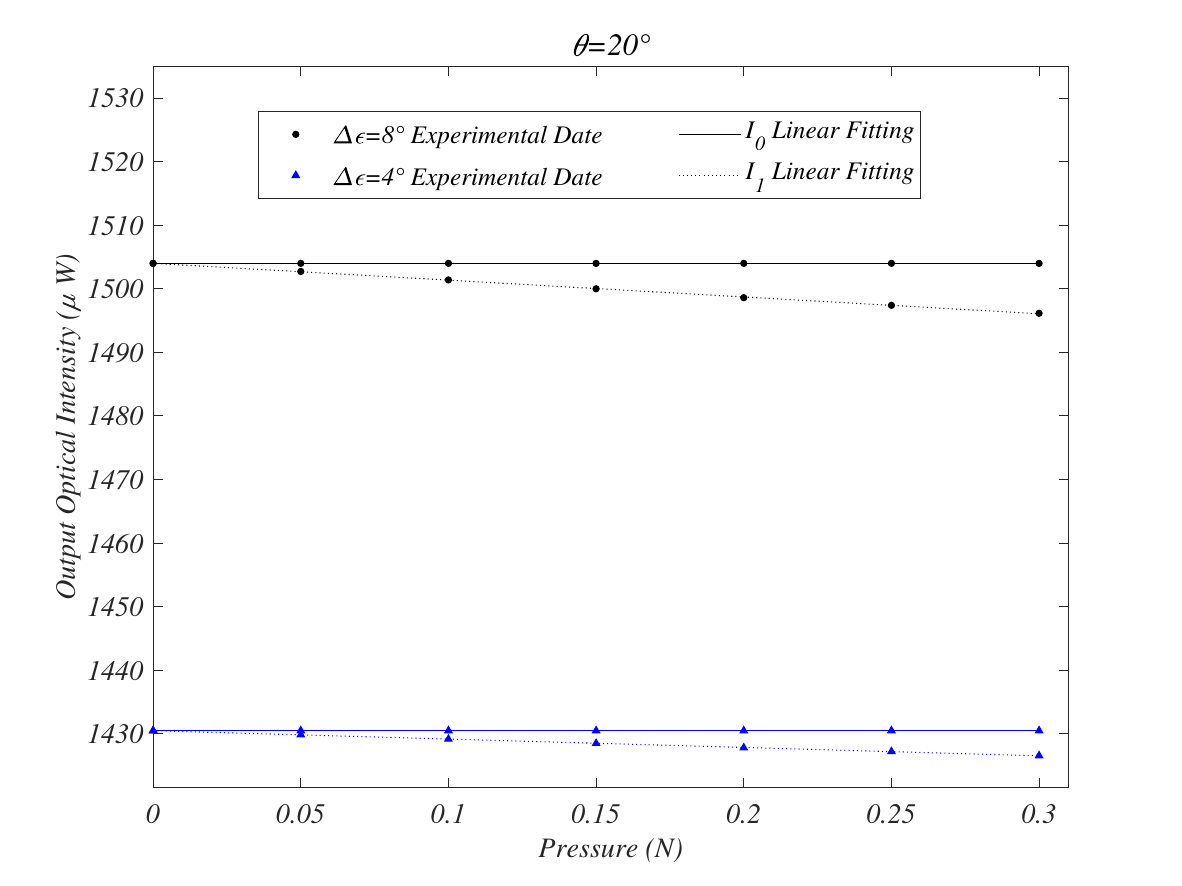}};
				\node at (0.40,3.0) {\textbf{(a)}};
			\end{tikzpicture}
		\end{minipage}
		\hfill
		\begin{minipage}{0.32\textwidth}
			\centering
			\begin{tikzpicture}
				\node[anchor=south west, inner sep=0] (image3) at (0,0) {\includegraphics[width=\linewidth]{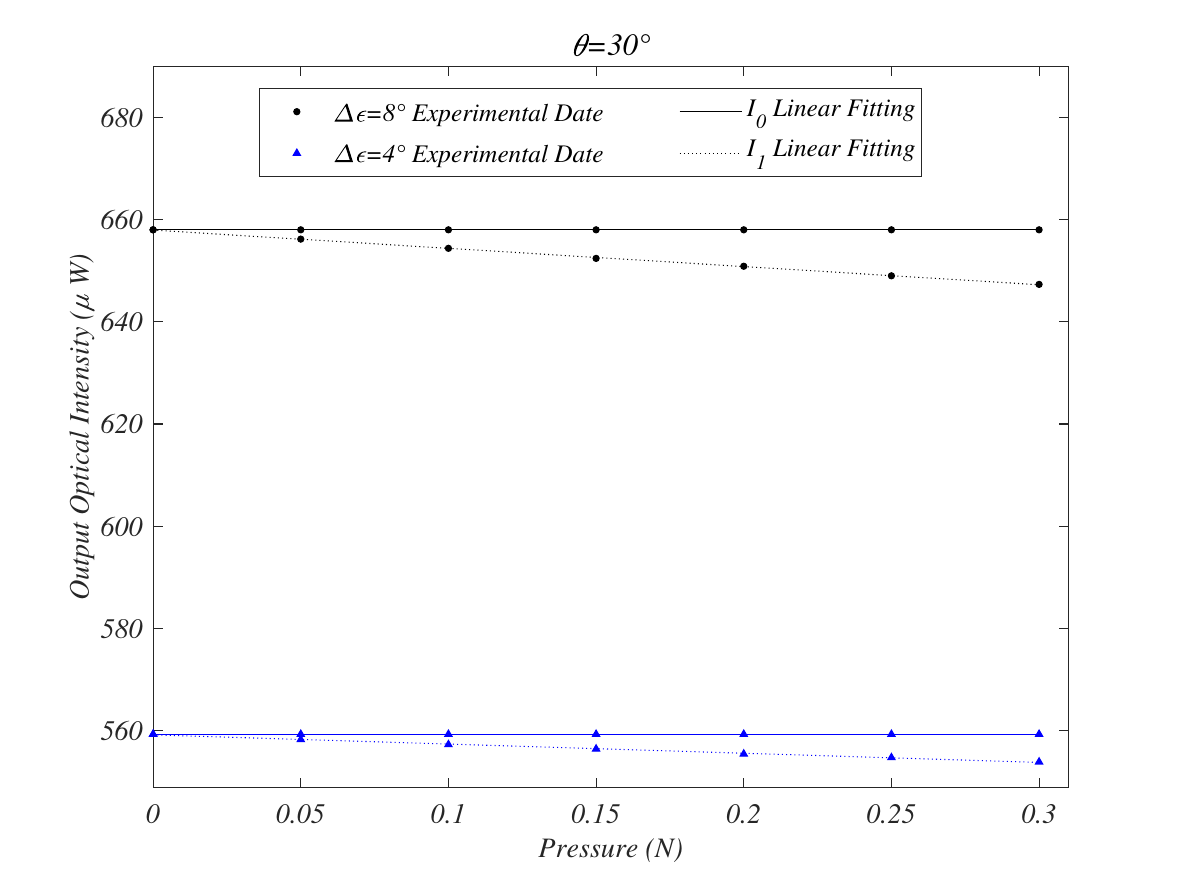}};
				\node at (0.40,3.0) {\textbf{(c)}};
			\end{tikzpicture}
		\end{minipage}
		\hfill
		\begin{minipage}{0.32\textwidth}
			\centering
			\begin{tikzpicture}
				\node[anchor=south west, inner sep=0] (image5) at (0,0) {\includegraphics[width=\linewidth]{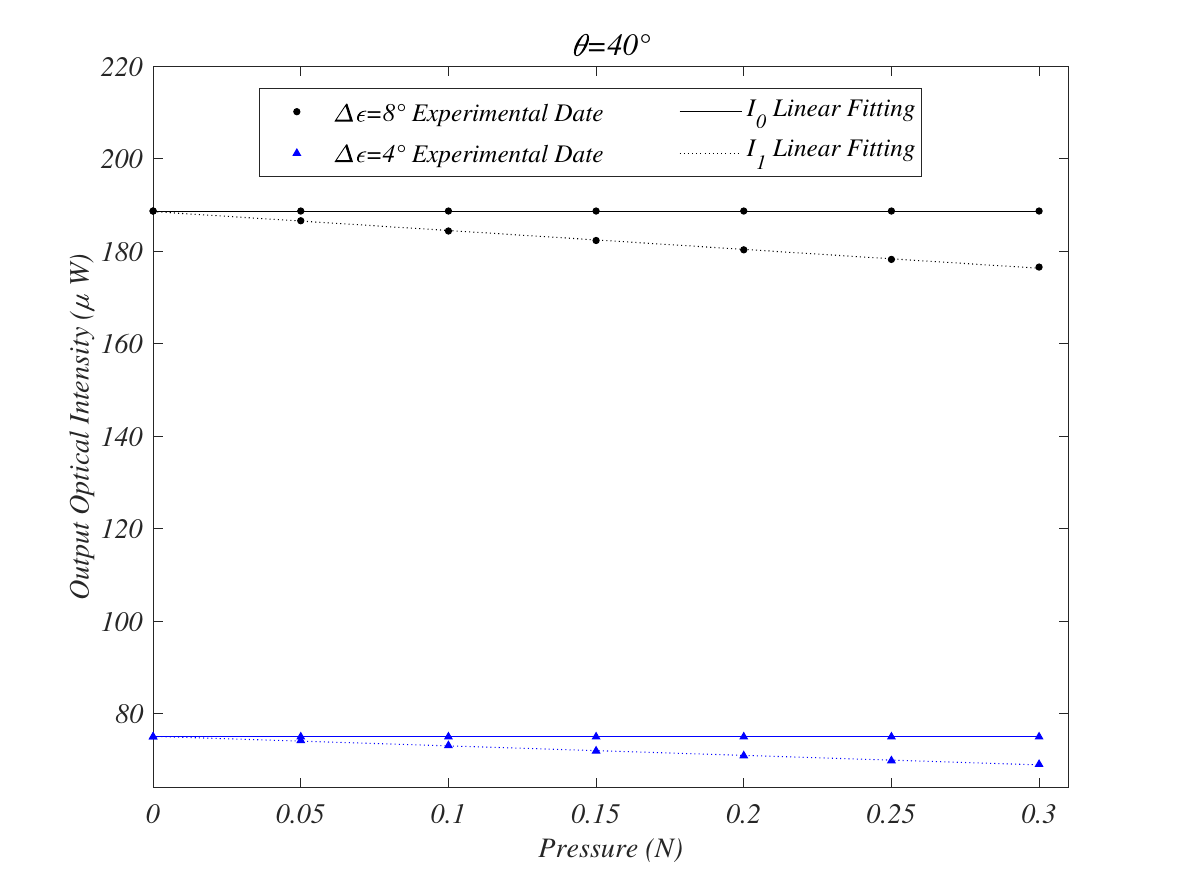}};
				\node at (0.40,3.0) {\textbf{(e)}};
			\end{tikzpicture}
		\end{minipage}
		
		\vspace{0cm}
		
		\begin{minipage}{0.32\textwidth}
			\centering
			\begin{tikzpicture}
				\node[anchor=south west, inner sep=0] (image2) at (0,0) {\includegraphics[width=\linewidth]{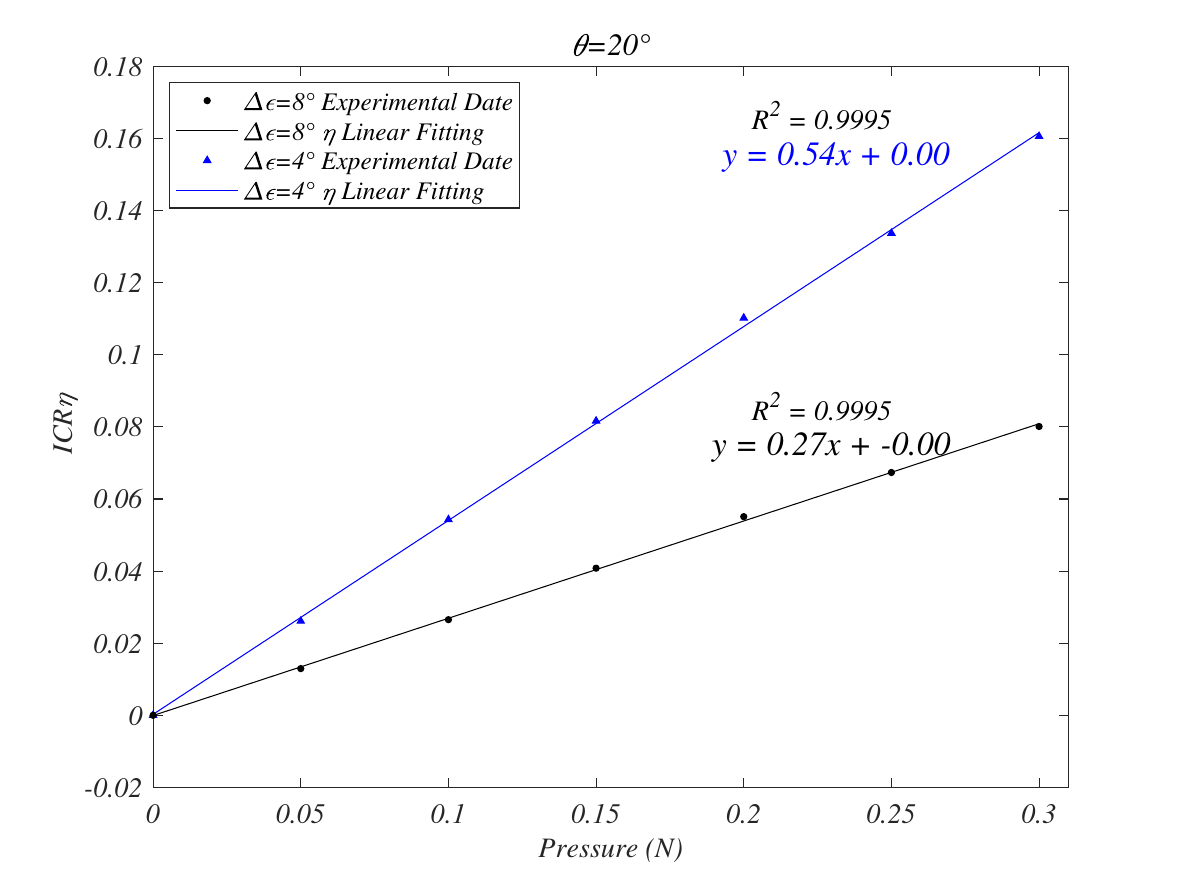}};
				\node at (0.40,3.0) {\textbf{(b)}};
			\end{tikzpicture}
		\end{minipage}
		\hfill
		\begin{minipage}{0.32\textwidth}
			\centering
			\begin{tikzpicture}
				\node[anchor=south west, inner sep=0] (image4) at (0,0) {\includegraphics[width=\linewidth]{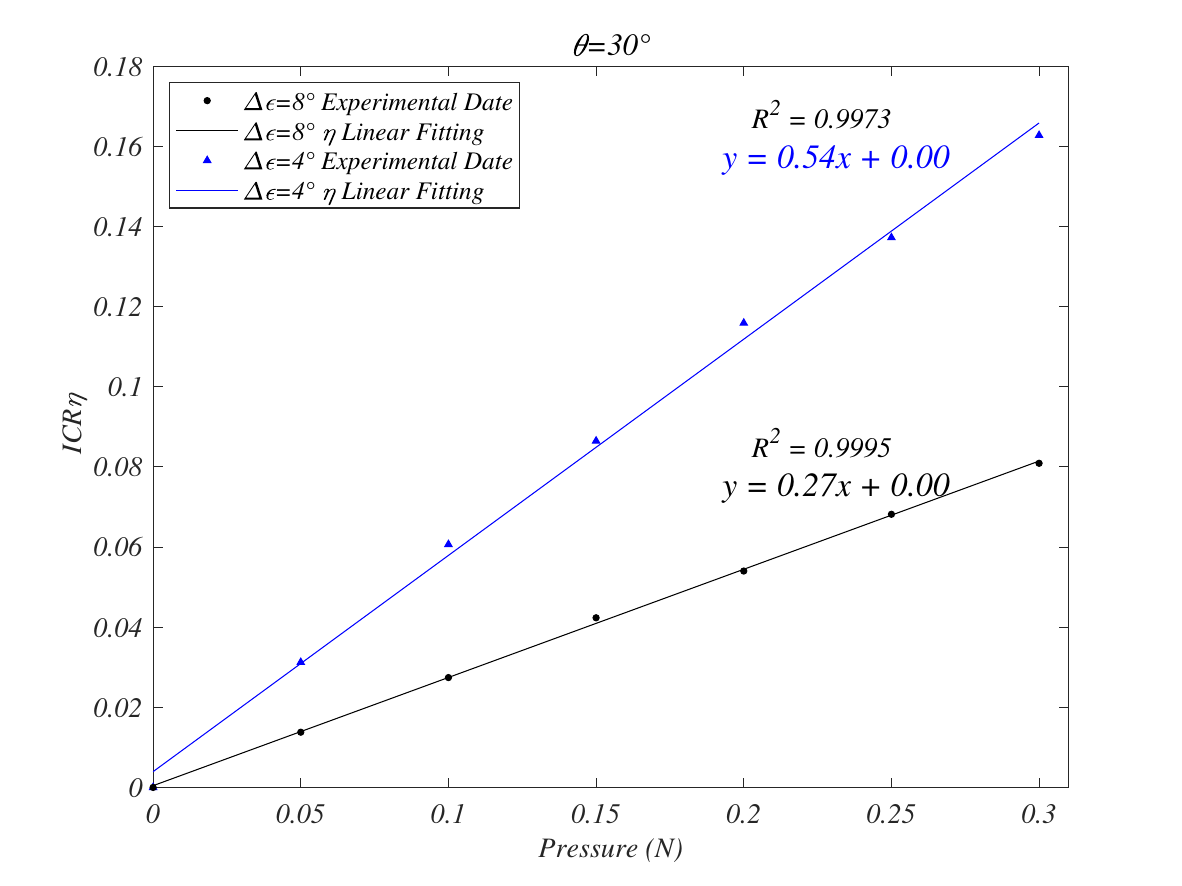}};
				\node at (0.40,3.0) {\textbf{(d)}};
			\end{tikzpicture}
		\end{minipage}
		\hfill
		\begin{minipage}{0.32\textwidth}
			\centering
			\begin{tikzpicture}
				\node[anchor=south west, inner sep=0] (image6) at (0,0) {\includegraphics[width=\linewidth]{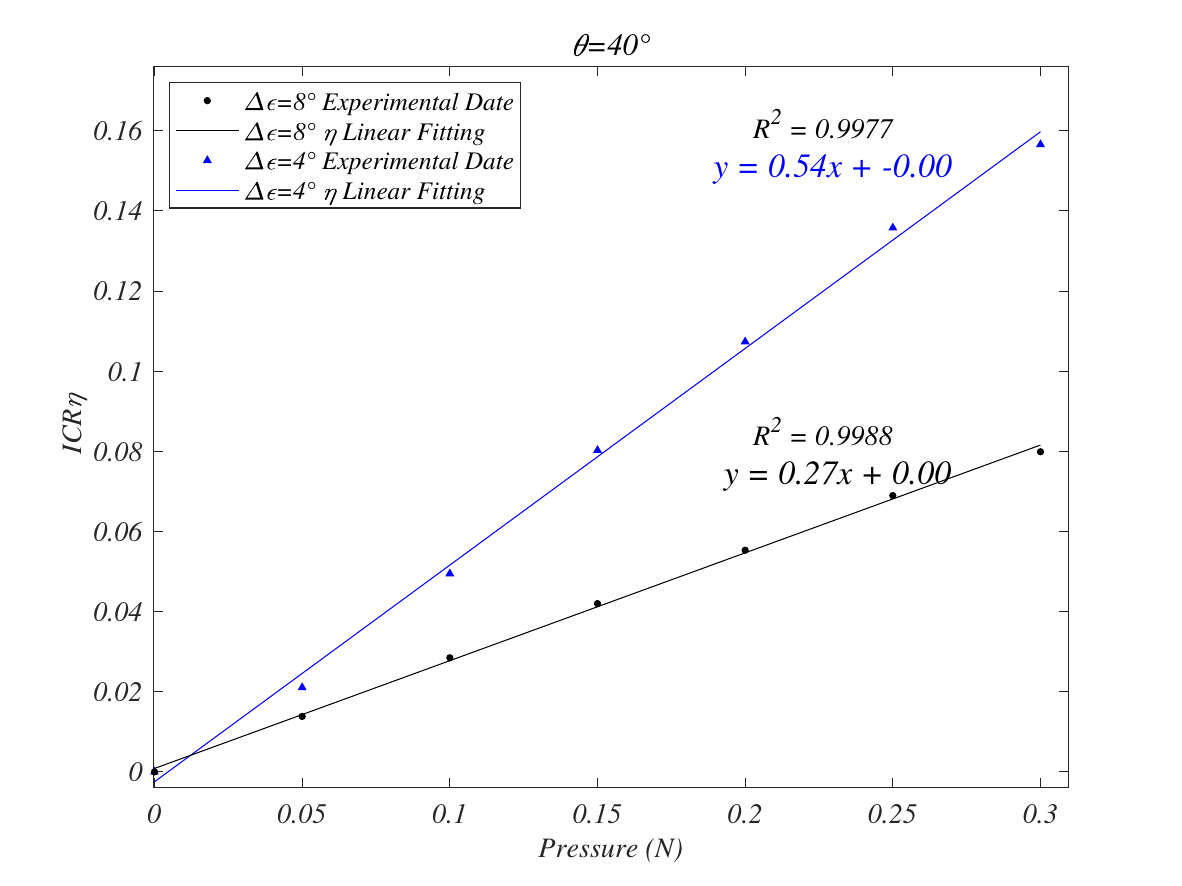}};
				\node at (0.40,3.0) {\textbf{(f)}};
			\end{tikzpicture}
		\end{minipage}
		
		\caption{Experimental results of pressure sensing with pre-selection angles $\theta = 20^{\circ}$, $30^{\circ}$, and $40^{\circ}$. Figures (a), (c), and (e) show the output optical intensity under optimized post-selection angles of $4^{\circ}$ and $8^{\circ}$ for each pre-selection setting. Figures (b), (d), and (f) depict the corresponding intensity contrast $\eta$ as a function of applied pressure for the same post-selection configurations}
		\label{fig:pressure_results}
	\end{figure}
	
	The experimental data demonstrate smaller $\Delta \epsilon$ yielding lower output intensities a trend consistent with theoretical predictions of post-selection probability. Fig. 4 (b), (d) and (f) show the dependence of intensity contrast $\eta$  on sensor pressure obtained by selecting different optimized post-selection angles under different pre-selection inputs. By linear fitting the experimental values, we obtained the strain sensitivity of the intensity contrast of $0.54/N$ and $0.27/N$ respectively when the pre-selected angle $\theta=20^{\circ}$ and the optimized post-selection angle $\Delta \epsilon$  was selected as 4° and 8° respectively. The linear fitting value $R^2$ was $0.9995$ and $0.9995$ respectively, indicating that the intensity contrast had a very good linearity to the pressure response.When expressed in units of intensity equivalent, the measured pressure sensitivity is $2.348dB/N$ and $1.174dB/N$ respectively.
	
	For pre-selection angles of $\theta=20^{\circ}$ and $40^{\circ}$, the optimized post-selection angles $\epsilon=4^{\circ}$ and $8^{\circ}$ yield intensity contrast pressure sensitivities of $0.54/N$ and $0.27/N$, corresponding to intensity-equivalent pressure sensitivities of $2.348 dB/N$ and $1.174 dB/N$ respectively. The pressure sensing experimental data strictly follow the linear relationship $\eta \approx 2\omega_{0} \Delta \tau /\Delta \epsilon$, in which the intensity contrast $\eta$ shows invariance to the pre-selected state of input and is inversely proportional to the optimized post-selection angle $\Delta \epsilon$, so that the pressure sensing sensitivity can be improved by reducing $\Delta \epsilon$.
	
	\subsection{Temperature sensing characteristics}
	The PMF itself has a thermo-optic coefficient, and the birefringence coefficient of PMF changes with temperature. When the ambient temperature of the sensing fiber PMF changes, the birefringence difference between the fast and slow axes of PMF also changes with temperature, introducing an optical phase difference between $H$ and $V$ polarizations, thus generating weak interactions. The phase difference produced by the temperature variation of PMF can be described by the following formula:
	\begin{equation}
		2\omega_{0} \Delta \tau =\frac{2\pi}{\lambda_{0}} \left( L\frac{\partial B}{\partial F} +B\frac{\partial L}{\partial F} \right) \Delta T,
	\end{equation} 
    
	where $B$  is the difference in refractive index between the fast and slow axes of the optical fiber, $L$ represents the PMF length, $\Delta T$  is the temperature change experienced by the PMF, and $\partial L/\partial T$  describes the fiber elongation rate caused by temperature changes in the PMF. When the sensing optical fiber PMF is only affected by temperature disturbance, the phase difference produced by the PMF can be considered proportional to the temperature change. The final relationship between the phase difference and the temperature change is: $2\omega_{0} \Delta \tau \sim 2\uppi \Delta T/\lambda_{0}$ .Thus, the change of phase difference of sensing optical fiber can be measured by WM sensing theory to reflect the change of temperature of sensing optical fiber.
	
	The experimental setup for preselection-free fiber-optic WM temperature sensing is shown in Fig. 5. During the experiment, a 30cm length of PMF was placed in a precision temperature chamber (temperature control accuracy: 0.01°C, thermal cycling test range from 5 to 80°C). The input and output fiber ports of the PMF in the precision temperature chamber share the same channel port, which is sealed with epoxy resin and tape to minimize heat leakage.
	\begin{figure}[htbp]
		\centering
		\includegraphics[width=1\linewidth]{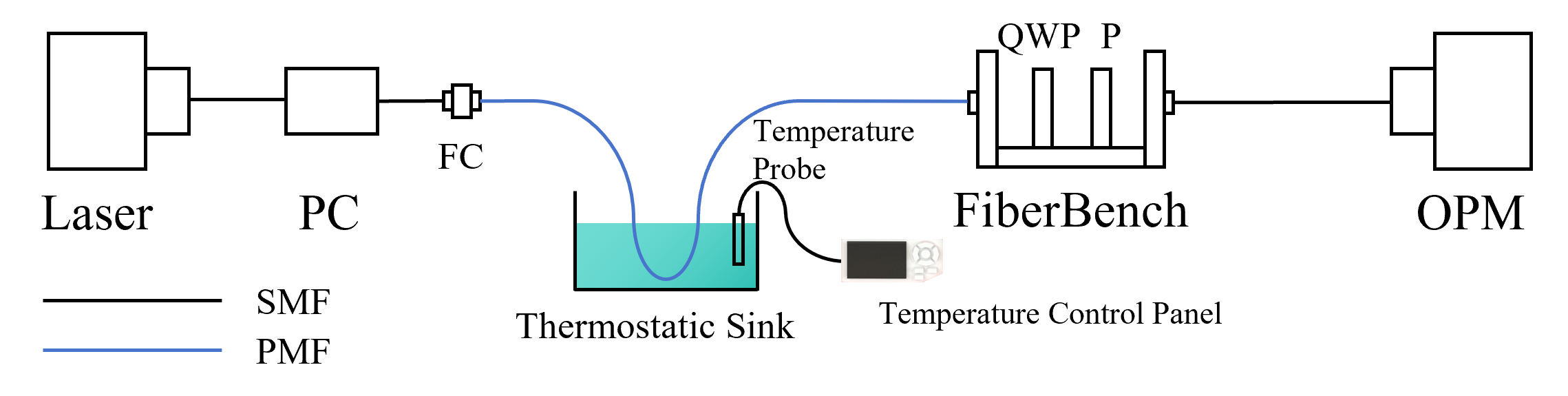}
		\caption{Schematic diagram of pre-selection-free fiber-optic WM temperature sensing}
		\label{fig:6}
	\end{figure}
	
	The experimental results in Figure 6 show the temperature sensing performance under three pre-selection states ($\theta=20^{\circ}$ [Fig. 6(a) and (b)], $30^{\circ}$ [Fig. 6(c) and (d)], $40^{\circ}$ [Fig. 6(e) and (f)]) at the optimized post-selection angles $\Delta \epsilon=8^{\circ}$ and 16°. Linear regression analysis revealed that as the temperature increased from 42.5°C to 42.8°C with a step change of 0.05°C, the output optical intensity exhibited a linear relationship with temperature. The corresponding linear fitting value $R^2$ were no less than 0.98, indicating a strong linear correlation between the output intensity and temperature variations.
	
	\begin{figure}[h]
		\centering
		
		\begin{minipage}{0.32\textwidth}
			\centering
			\begin{tikzpicture}
				\node[anchor=south west, inner sep=0] (image1) at (0,0) {\includegraphics[width=\linewidth]{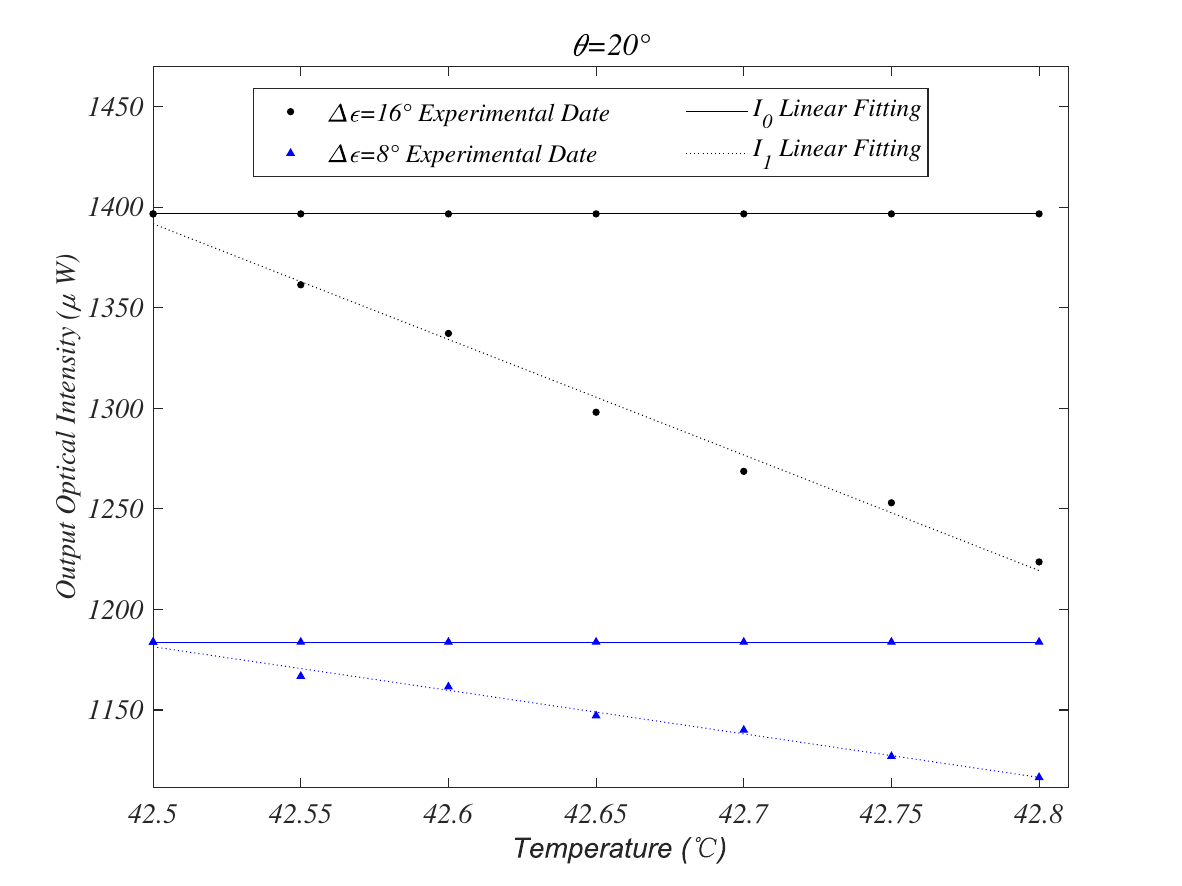}};
				\node at (0.40,3.0) {\textbf{(a)}};
			\end{tikzpicture}
		\end{minipage}
		\hfill
		\begin{minipage}{0.32\textwidth}
			\centering
			\begin{tikzpicture}
				\node[anchor=south west, inner sep=0] (image3) at (0,0) {\includegraphics[width=\linewidth]{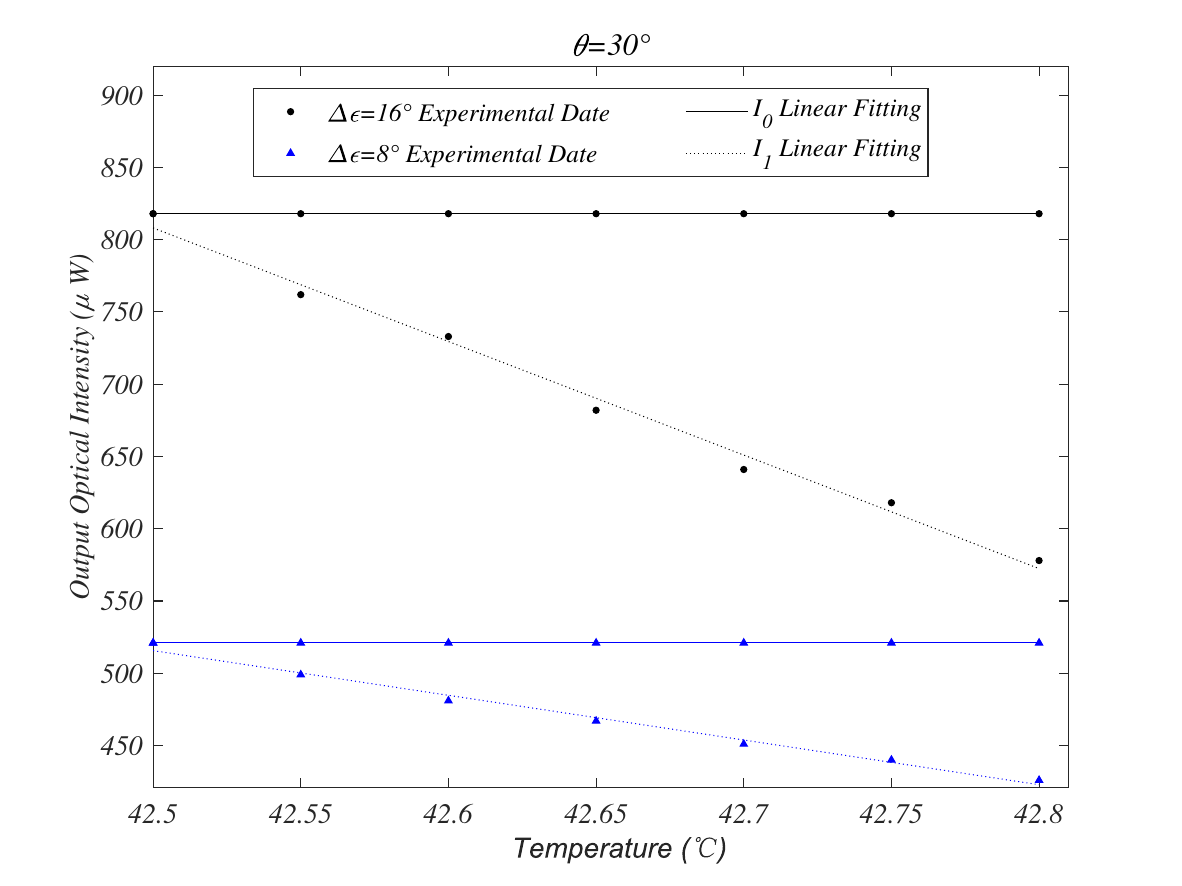}};
				\node at (0.40,3.0) {\textbf{(c)}};
			\end{tikzpicture}
		\end{minipage}
		\hfill
		\begin{minipage}{0.32\textwidth}
			\centering
			\begin{tikzpicture}
				\node[anchor=south west, inner sep=0] (image5) at (0,0) {\includegraphics[width=\linewidth]{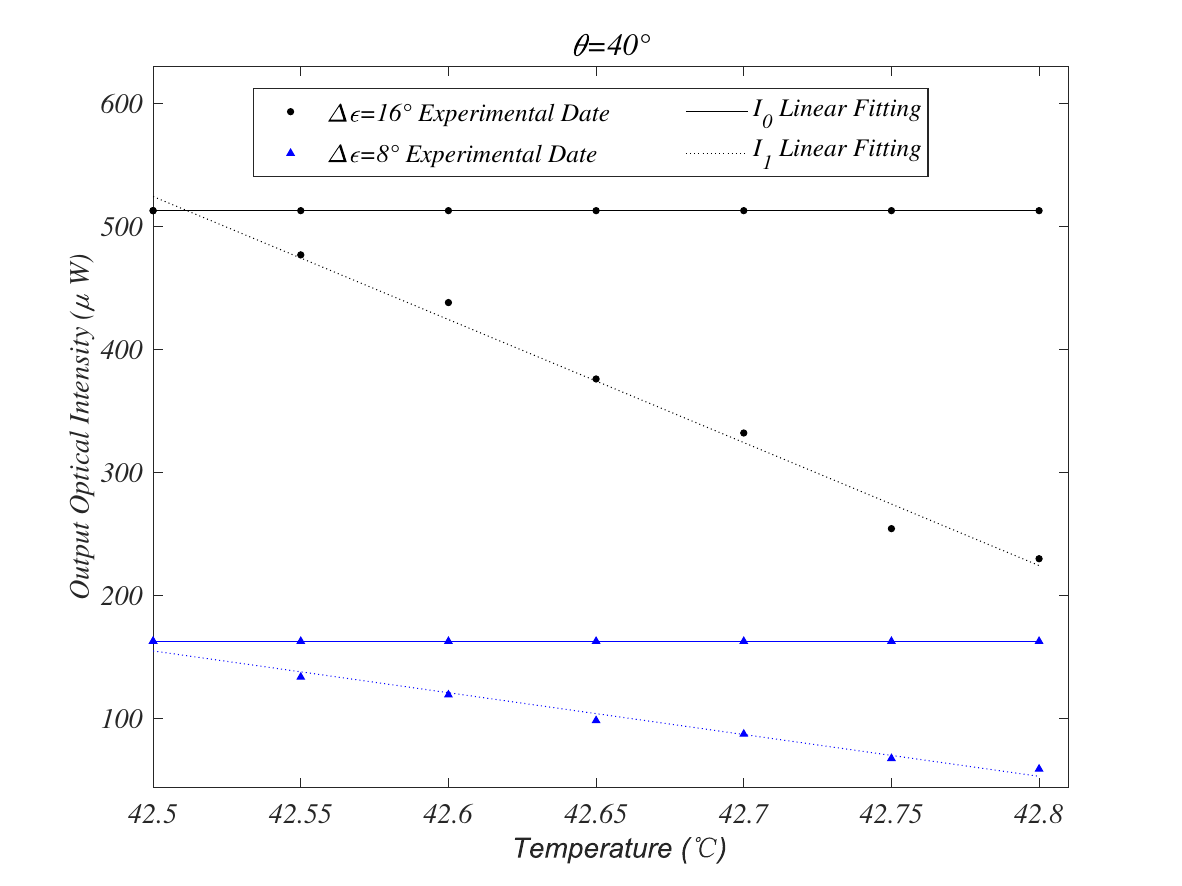}};
				\node at (0.40,3.0) {\textbf{(e)}};
			\end{tikzpicture}
		\end{minipage}
		
		\vspace{0.3cm}
		
		\begin{minipage}{0.32\textwidth}
			\centering
			\begin{tikzpicture}
				\node[anchor=south west, inner sep=0] (image2) at (0,0) {\includegraphics[width=\linewidth]{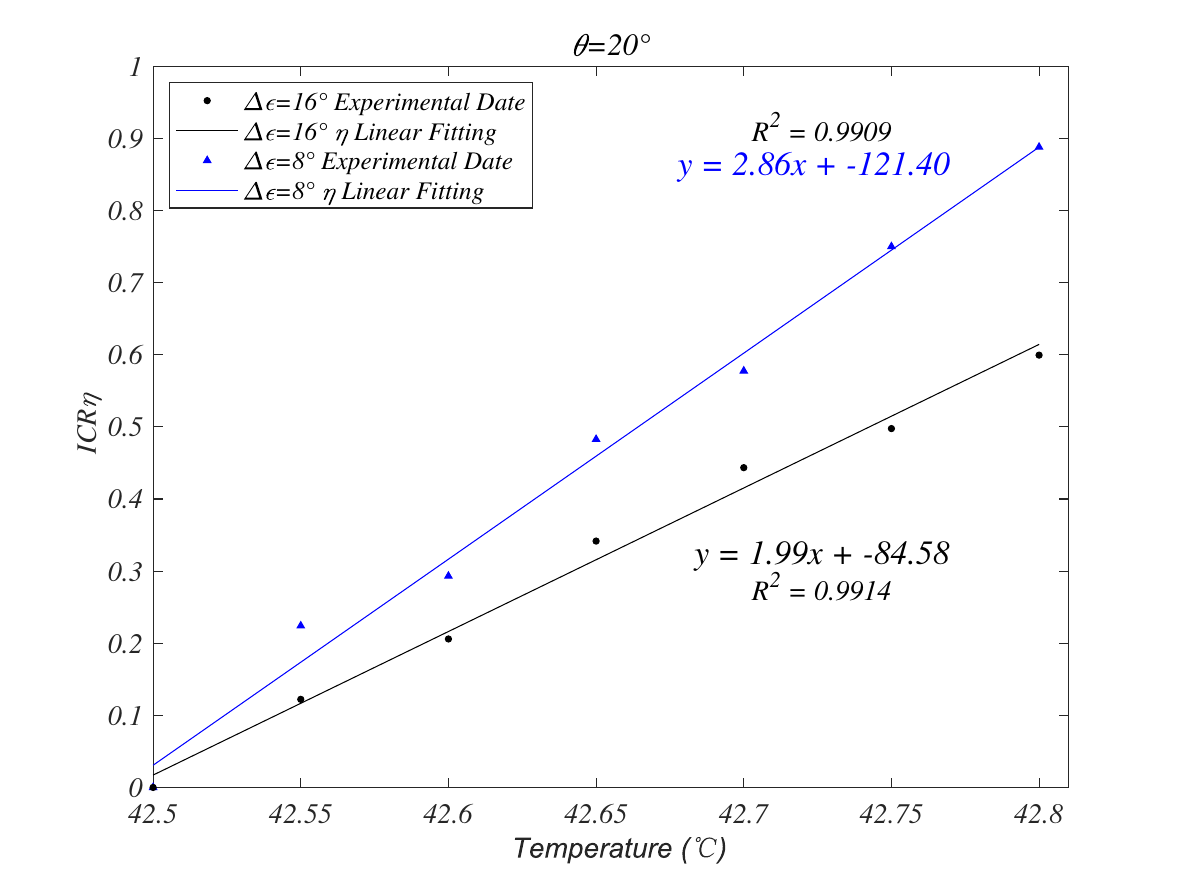}};
				\node at (0.40,3.0) {\textbf{(b)}};
			\end{tikzpicture}
		\end{minipage}
		\hfill
		\begin{minipage}{0.32\textwidth}
			\centering
			\begin{tikzpicture}
				\node[anchor=south west, inner sep=0] (image4) at (0,0) {\includegraphics[width=\linewidth]{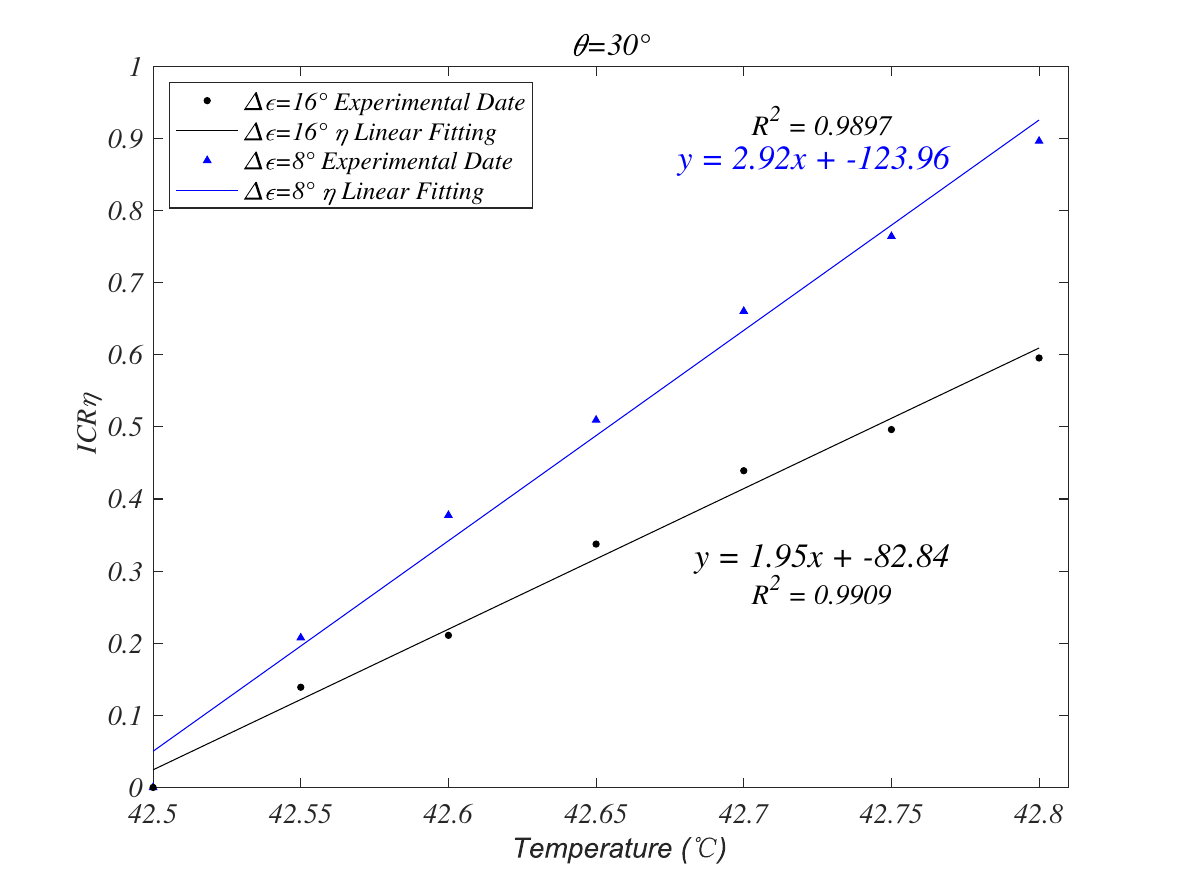}};
				\node at (0.40,3.0) {\textbf{(d)}};
			\end{tikzpicture}
		\end{minipage}
		\hfill
		\begin{minipage}{0.32\textwidth}
			\centering
			\begin{tikzpicture}
				\node[anchor=south west, inner sep=0] (image6) at (0,0) {\includegraphics[width=\linewidth]{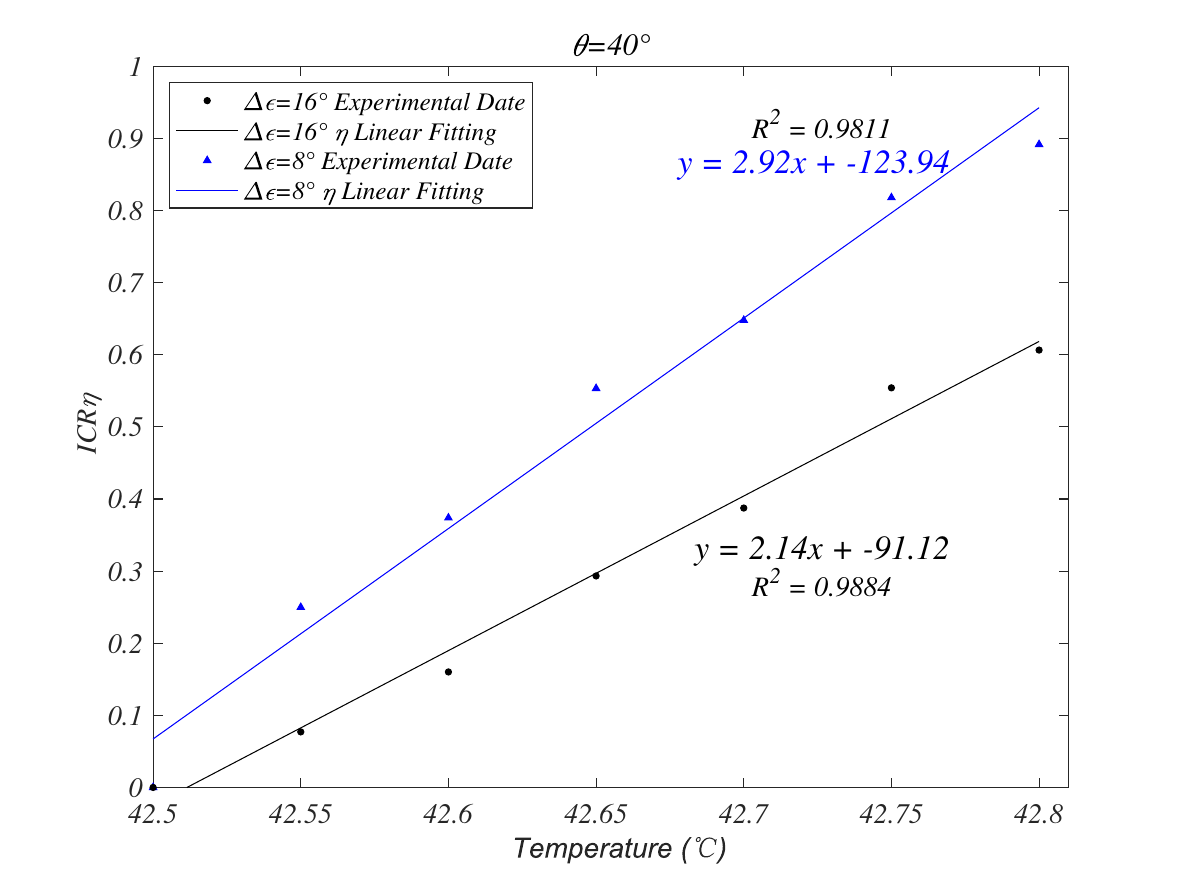}};
				\node at (0.40,3.0) {\textbf{(f)}};
			\end{tikzpicture}
		\end{minipage}
		
		\caption{Experimental results of temperature sensing for pre-selection angles $\theta = 20^{\circ}$, $30^{\circ}$, and $40^{\circ}$. Figures (a), (c), and (e) show the variation of output optical intensity with temperature. Figures (b), (d), and (f) present the corresponding intensity contrast $\eta$ versus temperature under optimized post-selection angles $\Delta \epsilon = 8^{\circ}$ (blue) and $16^{\circ}$ (black)}
		\label{fig:temperature_sensing}
	\end{figure}
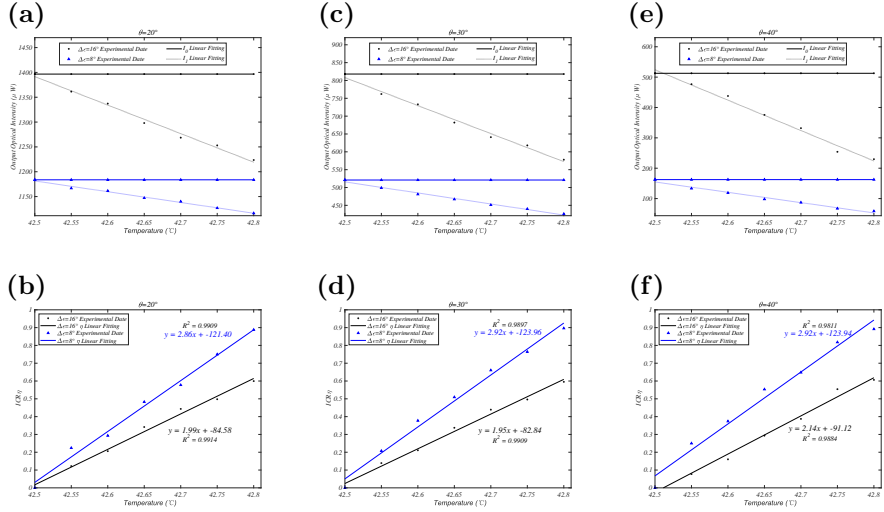
	
	Fig. 6 (b), (d) and (f) shows the relationship between intensity contrast  and sensing temperature for different pre-selection states ($\theta=20^{\circ}, 30^{\circ}$ and $40^{\circ}$) when the optimized post-selection angle $\Delta \epsilon$  is set to $8^{\circ}$ and $16^{\circ}$. The experimental values have been linearly fitted. When the pre-selection angle $\theta$ is set to $20^{\circ}$, and the optimized post-selection angles are $8^{\circ}$ and $16^{\circ}$, the temperature sensitivities derived from the experimental data are $2.86/^{\circ}C$ ($R^2=0.9909$) and $1.99/^{\circ}C$ ($R^2=0.9914$), respectively, with corresponding intensity equivalent temperature sensitivities of $12.435dB/°^{\circ}C$ and $8.652dB/^{\circ}C$. For the pre-selection angle $\theta$  of $30^{\circ}$, the temperature sensitivities are $2.92dB/^{\circ}C$ and $1.95dB/^{\circ}C$, with corresponding intensity equivalent temperature sensitivities of $12.695dB/°^{\circ}C$ and $8.478dB/^{\circ}C$, respectively. For the pre-selection angle $\theta$ of $40^{\circ}$, the temperature sensitivities are $2.92dB/^{\circ}C$ and $2.14dB/^{\circ}C$, with corresponding intensity equivalent temperature sensitivities of $12.695dB/^{\circ}C$ and $9.304dB/^{\circ}C$, respectively.
	
	The relationship between the intensity contrast $\eta$ obtained from different preselection angles and the sensitivity of the sensing temperature response is basically consistent, and the deviation can be attributed to the fluctuation of the experimental temperature. The temperature sensing sensitivity is only related to the optimized post-selection angle $\Delta \epsilon$ ; the smaller the optimized post-selection angle $\Delta \epsilon$ , the higher the intensity contrast temperature sensitivity obtained.
	
	\subsection{Analysis}
	The experimental results of the preselection-free fiber-optic weak measurement sensing schemethe to show ultra-sensitive response to physical parameter perturbations (phase, pressure, temperature),  achieving two to three orders of magnitude in sensitivity compared to traditional interferometric configurations (Table 1-2). Although grating-based sensing processes are relatively mature, their sensitivity is limited. In contrast, interferometric sensing schemes exhibit higher sensitivity. Various interferometer sensing schemes can use single or hybrid interferometric measurements. Interferometric configurations using bubble-shaped structured fibers or appropriate splicing for hybrid interferometric measurements may show higher sensitivity but are typically complex structures in manufacturing \cite{10508751}. Our sensing system employs ordinary PMF,  PC, and linear polarizer to achieve high-sensitivity phase sensing at $62 dB/rad$, high-sensitivity pressure sensing at $2.348 dB/N$, and high-sensitivity temperature sensing at $12.695dB/^{\circ}C$ without the need for spectral demodulation hardware. Crucially, the sensing sensitivity is only related to the selected optimized post-selection angle , not to the input pre-selected state.
	\begin{table}[htbp]
		\centering
		\caption{Comparing the sensitivity and complexity of water pressure sensing with classical sensing}
		\begin{tabular}{l c >{\hspace{-0cm}}c}  
			\toprule			\makecell{Sensor\\Configuration} &\makecell{Pressure\\Sensitivity}  &\makecell{Structural\\Complexity} \\
		\midrule
		DSH-FBG \cite{He:21} & 0.01398$nm/N$ & Moderate \\
		MZI and FPI\cite{WU2021166962}&1.596$nm/N$                        & Complex \\  
				HB-SCF and SI \cite{9720981}                                      & 0.28837$nm/N$                     & Simple  \\  
				PMF FBG and SI \cite{ZHANG2023103423}                  & 408.7$pm/N$                        & Moderate \\  
				SMF \cite{10508751}                                                       &  90.3$nm/N$                         & Complex \\  
				PCF and Si \cite{8374857}                                              & 0.6201$nm/N$                      & Simple \\  
				PMF (This work)                                                              & 2.348$dB/N$                          & Simple \\  
				\bottomrule
    \end{tabular}
    \end{table}

	\begin{table}[htbp]
		\centering
		\caption{Comparing the sensitivity and complexity of temperature sensing with classical sensing}
		\begin{tabular}{l c >{\hspace{-0cm}}c}  
			\toprule
			\makecell{Sensor\\Configuration} &\makecell{Temperature\\Sensitivity}  & \makecell{Structural\\Complexity} \\
			\midrule
			Cascade FBG \cite{10.111}                                            & 10$pm/^{\circ}C$                                   & Moderate  \\  
			Microfiber Junction
			\cite{s1410}                   
			 & 22.81$pm/^{\circ}C$                              & Moderate \\  
			FPI \cite{Liu:15}                                                              & 84.6$pm/^{\circ}C$                                & Moderate  \\  
			C-PCF \cite{WU20}                                                        & 1.054$nm/^{\circ}C$                               & Complex \\  
			PDMS-SNCS \cite{WANG}                                            & 260$pm/^{\circ}C$                                   & Moderate \\  
			SMF-MMF-HCF-\\HCF-MMF-SMF \cite{ZHAO}               & 279.99$pm/^{\circ}C$                             & Moderate \\  
			Cascade FPI and MZI \cite{YUAN}                                 & 305.42$pm/^{\circ}C$                             & Simple \\  
			PMF (This work)                                                             & 12.695$dB/^{\circ}C$                                & Simple \\  
				\bottomrule
    \end{tabular}
    \end{table}
	
	\section{Conclusion}
	In this study, we propose and experimentally demonstrate a preselection-free fiber-optic weak measurement sensing framework. Compared with previous fiber-optic weak measurement sensing structures, this preselection-free scheme is simpler, and high-sensitivity sensing can be achieved by adjusting the optimized post-selection angle. Experimental results show that this fiber optic WM sensing scheme can achieve nearly consistent high-sensitivity performance with different pre-selections, achieving a phase sensitivity of up to $62 dB/rad$ and a phase resolution of $1.6\times10^{-5} rad$; a pressure sensitivity of $2.348 dB/N$ and a pressure resolution of $4.2\times10^{-4} N$; a temperature sensitivity of $12.695dB/^{\circ}C$ and a temperature resolution of $7.8\times10^{-5}\ ^{\circ}C$. 
	
	Due to various disturbances and noise in actual experiments, there will be some error in the detected light intensity, resulting in the actual sensing accuracy being one or two orders of magnitude lower than the theoretical accuracy. The core advantage of this sensing scheme lies in its simple structure (only requiring PMF, PC, and a polarizer), no need for complex demodulation equipment, and flexible active sensitivity adjustment through the optimized post-selection angle, which can achieve two to three orders of magnitude higher sensitivity compared to traditional fiber optic sensing measurements.
	
	Compared to existing fiber Bragg grating or fiber-optic WM technologies, the manufacturing cost is significantly reduced, and it avoids the impact of pre-selection state preparation on system stability. This provides new solutions for industrial monitoring (such as microforce detection and pipeline leak warnings) and biomedical applications (such as cell mechanical characterization). Future work will focus on developing multi-parameter decoupling algorithms and long-term stability testing to further expand its application in distributed sensor networks.
	
	\bibliographystyle{unsrt}  
	\bibliography{main}

\end{document}